\documentstyle [11pt,epsf,rotate] {article}

\def\Tr{\mbox{Tr}\,}
\def\p{\tilde{p}}
\def\q{\tilde{q}}
\def\ds{\displaystyle}
\def\J{\mbox{J}}

\begin{document}

\noindent ULM--TP/97-1 \\
January 1997
\vspace{3.0cm}

\centerline{\huge Bifurcations of Periodic Orbits}
\vspace{0.5cm}
\centerline{\huge and Uniform Approximations}
\vspace{1.0cm}
\centerline{\Large Henning Schomerus$^1$ and Martin Sieber$^2$}
\vspace{1.2 cm}

\noindent $^1$ Fachbereich Physik, Universit\"at-Gesamthochschule
Essen, D-45117 Essen, Germany \\
\noindent $^2$ Abteilung Theoretische Physik, 
Universit\"at Ulm, D-89069 Ulm, Germany

\vspace{2.0cm}
\centerline{\bf Abstract}
\vspace{0.5cm}

We derive uniform approximations for contributions to Gutzwiller's 
periodic-orbit sum for the spectral density which are valid close 
to bifurcations of periodic orbits in systems with mixed
phase space. 
There, orbits lie close together and give collective contributions,
while the individual contributions of Gutzwiller's type would diverge
at the bifurcation. New results for the tangent, the period
doubling and the period tripling bifurcation are
given.  They are obtained by going beyond the local approximation 
and including higher order terms in the normal form of the action.
The uniform approximations obtained are tested on the kicked top and
are found to be in excellent agreement with exact quantum results.

\vspace{2.5cm}

\noindent PACS numbers: \\
\noindent 03.20.+i ~ Classical mechanics of discrete systems: 
general mathematical aspects. \\
\noindent 03.65.Sq ~ Semiclassical theories and applications. \\
\noindent 05.45.+b ~ Theory and models of chaotic systems.

\vspace{0.3cm}
\noindent{\it Submitted to Journal of Physics A}

\newpage

\section{Introduction}
\label{secintro}

Semiclassical approximations in terms of periodic orbits
belong to the main tools for the examination of spectral
properties of quantum systems. They allow, for example, to 
explain fluctuations in quantum
spectra in terms of the periodic orbits of the
corresponding classical system. Semiclassical
periodic-orbit approximations have been derived in
cases where the classical dynamics is chaotic or
integrable or has more general symmetries
\cite{Gut71,BB72,BB74,BT76,BT77a,Gut90,CL91,CL92}.
In these systems periodic orbits are typically either
isolated or appear in families.

Most systems, however, are neither chaotic nor integrable
but show a complicated mixture of regular and chaotic behaviour.
In these systems semiclassical approximations are more
complicated and up to date there do not exist complete
semiclassical approximations for quantities like the
spectral density in terms of periodic orbits.
These difficulties are due to the fact that in many
situations periodic orbits neither appear in families
nor can they be treated as being isolated in a semiclassical 
approximation. This is the case when there are other
periodic orbits closely nearby. 
If the action differences of neighbouring periodic orbits
is not large in comparison to $\hbar$ then the orbits yield
a collective semiclassical contribution, and this is the 
typical situation when bifurcations
of periodic orbits occur. In order to extend semiclassical
approximations to systems with mixed phase space one has to
derive expressions for the joint contribution of orbits
which participate in a bifurcation. For generic two-dimensional
systems this was first done by Ozorio de Almeida and Hannay. They
derived an approximation which is valid in the vicinity
of a bifurcation \cite{OH87}. In the present paper we extend
the results of Ozorio de Almeida and Hannay and derive analytic formulas
which interpolate over the regime from a bifurcation
up to regions where the orbits can be considered isolated.

Bifurcations are a characteristic phenomenon in systems
with mixed phase space. They are responsible for the
rapid increase of the number of periodic orbits when an
integrable system is transformed into a chaotic system,
e.\,g.\ by changing an external parameter. If one changes
this parameter by an arbitrarily small but finite amount,
then in general an infinite number of bifurcations occur,
since they take place any time that the stability angle
of a stable orbit is a rational multiple of $2 \pi$.
There are different kinds of generic bifurcations,
but the number of different forms is limited.
The generic bifurcations that occur in two-dimensional
conservative systems (or, equivalently, one-dimensional
area-preserving maps) were classified by Meyer and Bruno 
\cite{Mey70,Brj70,Bru72}. 
They are characterized by normal forms which
describe the characteristic classical motion in the 
vicinity of a periodic orbit. 
Altogether one has to distinguish five different
cases. These are the period-$m$-tupling bifurcations with
$m=1$ up to $m=5$. They have the property that a central
periodic orbit bifurcates and other periodic orbits
split from the central orbit whose primitive period is $m$
times the primitive period of the central orbit. (An
exception is the case $m=1$ for which there is
no periodic orbit before the bifurcation.)
The cases
for $m>5$ follow the same pattern as for $m=5$.

Ozorio de Almeida and Hannay derived their approximation
by an integration over the normal forms. Their results are
expressed in terms of diffraction catastrophe integrals.
We call this approximation in the following the local
approximation, since it is valid in the vicinity of a
bifurcation. In farther distance from a bifurcation it
reduces to an approximation of Gutzwiller's type for
isolated periodic orbits, it does not yield, however,
the correct semiclassical amplitudes of the orbits.
In \cite{Sie96} the results of Ozorio de Almeida were
extended for the case $m \geq 5$ by including 
higher order terms in the normal forms. A slightly
generalized version of this result is given in \cite{Sie96b}.
In the present
paper we derive corrections for the low order bifurcations
with $m=1$ up to $m=3$. We obtain uniform
approximations in closed form which are valid from a bifurcation up to
the regime where Gutzwiller's approximation can be applied.

The article is organized as follows. In section 2 we
introduce the semiclassical method for treating 
bifurcations and in section 3 we present the results
for the uniform approximations in two-dimensional
conservative systems that are derived in the appendices.
In section 4 we give the corresponding results for
one-dimensional area-preserving maps. Numerical
examinations of the uniform approximations are
carried out in section 5 on the example of a kicked
top, and in section 6 we discuss the range of validity
of our approximations and possible further extensions.

\section{Contributions to the spectral density}
\label{secdofe}

In autonomous systems with discrete energy spectra
the density of states
\begin{equation} \label{sec2a}
d(E) = \sum_n \delta(E - E_n) = - \frac{1}{\pi} \mbox{\bf Im\,} \Tr G(E)
\end{equation}
can be expressed in terms of the trace of the (retarded)
Green function.

Semiclassical contributions of periodic orbits to the
level density are derived from (\ref{sec2a}) by inserting
the semiclassical approximation for the Green function. In
a mixed coordinate-momentum representation this approximation
is given for two-dimensional systems by
\begin{equation} \label{sec2b}
G(\vec{q}\,', \vec{p}, E) \approx
\frac{1}{i \hbar \sqrt{2 \pi i \hbar}} 
\sum_\xi \sqrt{|D_\xi|}
\exp \{ \frac{i}{\hbar} \hat{S}_\xi(\vec{q}\,',\vec{p},E)
-\frac{i \pi}{2} \hat{\nu}_\xi \} \; ,
\end{equation}
where the sum runs over all classical trajectories with initial
momentum $\vec{p}$ and final position $\vec{q}\,'$ at energy $E$. 
The function $\hat{S}_\xi$ for a trajectory $\xi$ is defined as 
\begin{equation} \label{sec2c}
\hat{S}_\xi(\vec{q}\,',\vec{p},E) = \int_{\vec{q}}^{\vec{q}\,'} 
\! \vec{p} \cdot \mbox{d} \vec{q} +  \vec{q} \cdot \vec{p} \; ,
\end{equation}
where the integral is evaluated along the trajectory $\xi$,
and $\vec{q}$ is the initial position of the trajectory
which is determined by the initial momentum $\vec{p}$ and
the final position $\vec{q}\,'$. Its value follows from
the condition that the gradient of the right-hand side of
(\ref{sec2c}) with respect to $\vec{q}$ vanishes. 
$\hat{S}(\vec{q}\,',\vec{p},E)$ is the generating function
for a canonical transformation from final to initial coordinates
of a trajectory. The mixed representation has been chosen
since for a bifurcating orbit this transformation is close
to the identity which cannot be generated in a pure
coordinate- or momentum-representation.

Finally, $D_\xi$ is the determinant of a matrix of second derivatives
of $\hat{S}_\xi$
\begin{equation} \label{sec2d}
D_\xi = \det \left( \begin{array}{cc}
\frac{\partial^2 \hat{S}_\xi}{\partial \vec{q}\,' \partial \vec{p}} &
\frac{\partial^2 \hat{S}_\xi}{\partial \vec{q}\,' \partial E} \\ [4pt]
\frac{\partial^2 \hat{S}_\xi}{\partial E \partial \vec{p}} &
\frac{\partial^2 \hat{S}_\xi}{\partial E^2}
\end{array} \right) \; ,
\end{equation}
and $\hat{\nu}_\xi$ is an integer related to the Maslov index
(see e.\,g.\ \cite{CL91}).

The semiclassical spectral density is then determined by
\begin{equation} \label{sec2e}
d(E) = - \frac{1}{\pi} \, \mbox{\bf Im} \int \! \mbox{d}^2 q\,'
\, \mbox{d}^2 p \, G(\vec{q}\,', \vec{p}, E) \, 
\exp \left( - \frac{i}{\hbar} \vec{q}\,' \cdot \vec{p} \right) \; .
\end{equation}

In the vicinity of a periodic orbit the integrals are evaluated
by choosing local coordinates with one coordinate along the
orbit and one perpendicular to it. If the integral over
the perpendicular direction is evaluated in stationary
phase approximation one obtains the Gutzwiller approximation
for the semiclassical contribution of the orbit.
\begin{equation} \label{sec2ee}
d_\xi(E) = \frac{1}{\pi \hbar} 
\frac{T_\xi}{r_\xi \sqrt{|\mbox{Tr\,} M_\xi - 2|}}
\cos \left( \frac{S_\xi}{\hbar} - \frac{\pi}{2} \nu_\xi \right) \; .
\end{equation}
Here $S_\xi$ is the action of the orbit whose value is given by
the function $\hat{S}_\xi(\vec{q}\,',\vec{p},E) - \vec{q}\,' \cdot \vec{p}$
for the considered periodic orbit. $T_\xi$, $M_\xi$ and $\nu_\xi$
are the period, stability matrix and Maslov index of the orbit,
respectively. The integer $r_\xi$ denotes the repetition
number of the orbit. In this notation we consider a periodic orbit
and its multiple traversals which all give semiclassical
contributions to the level density as different periodic orbits.
The uniform approximations for bifurcating periodic orbits that
are derived in this paper will be expressed in terms of exactly
the same classical quantities that appear in (\ref{sec2ee}).

We consider now the contribution of a bifurcating orbit with
repetition number $r$ to the level density. The condition for
such a bifurcation is that the stability matrix of the
corresponding primitive periodic orbit (repetition number $r=1$) 
satisfies $\mbox{Tr\,} M = 2 \cos (2 \pi n/r)$
with integer $r$ and $n$ and thus the stability matrix of 
the $r$-th traversal has a trace which is equal to two.
Let $l$ be the greatest common divisor of $r$ and $n$. 
Then the bifurcation is a period-$m$-tupling bifurcation
with $m=r/l$.
Near the bifurcation the integral over the perpendicular
coordinates cannot be evaluated in stationary phase approximation
since there are stationary points nearby which correspond to the
other orbits participating in the bifurcation. This is
reflected by the fact that the Gutzwiller approximation
(\ref{sec2ee}) diverges at the bifurcation.
Instead one has to derive a joint contribution
of all orbits which are involved in the bifurcation.
This is achieved by expanding the generating function
$\hat{S}(\vec{q}\,',\vec{p},E)$ in higher order around the
central orbit. In general this results in a complicated
exponent in the integrand of (\ref{sec2e}). 
The integrals can be considerable simplified by a canonical
transformation of the coordinates and by using the fact that
the form of eq.\,(\ref{sec2e}) is semiclassically invariant
under canonical transformations. This follows from work of
Miller \cite{Mil74} and is discussed by Littlejohn \cite{Lit90}.

The most simple form that the generating function can take
near the bifurcation is given by the normal form. This
normal form contains the information about the number
and arrangement of the orbits which are involved in 
the bifurcation. The transformation to the normal
form coordinates has the further advantage that
then the integral over the coordinate along the 
periodic orbit can be performed trivially. These
steps are explained in detail in \cite{Sie96} and for
that reason we give here only the resulting formula
\begin{equation} \label{sec2f}
d_\xi(E) \approx \frac{1}{2 \pi^2 \hbar^2} \, \mbox{\bf Re} 
\int_{-\infty}^\infty \! \mbox{d}q' \, 
\int_{-\infty}^\infty \! \mbox{d}p  \, \frac{1}{r} \,
\frac{\partial \hat{S}}{\partial E} \,
\left| \frac{\partial^2 \hat{S}}{\partial p \partial q'}
\right|^{\frac{1}{2}}
\exp \{ \frac{i}{\hbar} \hat{S}(q', p, E) 
- \frac{i}{\hbar} q' p - \frac{i \pi}{2} \nu \} \; .
\end{equation}
Here $p$ and $q'$ are coordinates in the Poincar\'e
surface of section perpendicular to the orbit,
and $\hat{S}(q', p, E)$ is the generating function 
for the $r$-th iterate of the Poincar\'e map
which obeys the conditions
\begin{equation} \label{sec2g}
\frac{\partial \hat{S}}{\partial q'} = p'
\; , \; \; \;
\frac{\partial \hat{S}}{\partial p}  = q
\; , \; \; \;
\frac{\partial \hat{S}}{\partial E} = T \; ,
\end{equation}
where $T$ is the time from initial to final point.

The approximation of Ozorio de Almeida and Hannay
for the contributions of orbits near a bifurcation
is obtained by inserting the normal form of the
generating function $\hat{S}(q', p, E)$ for a particular
generic bifurcation. This yields the semiclassical
contributions in terms of standard diffraction
catastrophe integrals. For the 
bifurcations which are considered in this paper
the normal forms are given by
\begin{equation} \label{sec2h} \begin{array}{ll}
m = 1: \; \; \; \; \; \; \; \; \; \; &
\hat{S}(q', p, E) = S_0(E) + q' p 
- \frac{\ds \sigma}{\ds 2} p^2 - \varepsilon q' - a q'^3 \\[6pt]
m = 2: &
\hat{S}(q', p, E) = S_0(E) + q' p 
- \frac{\ds \sigma}{\ds 2} p^2 - \varepsilon q'^2 - a q'^4 \\[6pt]
m = 3: &
\hat{S}(q', p, E) = S_0(E) + q' p 
- \frac{\ds \varepsilon}{\ds 2}(q'^2 + p^2) 
- \frac{\ds a}{\ds \sqrt{8}} (p^3 - 3 p q'^2) \; ,
\end{array}
\end{equation}
where $\varepsilon$ is a parameter which is zero
at the bifurcation and $\sigma$ is a sign factor. 
As will be shown in the appendices, the diffraction
integrals for these cases can be expressed
in terms of Bessel functions. The properties
of the bifurcations corresponding to these
normal forms are discussed in more detail in the
next section. All these bifurcations involve
two periodic orbits. 

The approximation of Ozorio de Almeida and Hannay is
valid in the vicinity of a bifurcation and hence is called local
approximation by us.
In farther distance from a
bifurcation the local approximation splits up asymptotically into a
sum of separate contributions of Gutzwiller's type
for the periodic orbits. However, in this limit
the semiclassical amplitudes of the orbits have a
fixed ratio, i.\,e.\ there is a fixed relationship
between the stabilities (and periods) of the different
orbits. In more detail, the approximation holds when
the following relations between the monodromy matrices
and the periods of the orbits are valid.
\begin{equation} \label{sec2i} \begin{array}{lll}
m = 1: \; \; \; \; \; \; \; \; \; \; &
\mbox{Tr\,} M_1 + \mbox{Tr\,} M_2 - 4 = 0 \, ,
\; \; \; \; \; \; & T_1 = T_2 \\[4pt]
m = 2: & 
\mbox{Tr\,} M_1 + 2 \, \mbox{Tr\,} M_0 - 6 = 0 \, , 
& T_1 = T_0 \\[4pt]
m = 3: &
\mbox{Tr\,} M_1 + 3 \, \mbox{Tr\,} M_0 - 8 = 0 \, ,
& T_1 = T_0 \; .
\end{array}
\end{equation}
For $m=2$ and $m=3$ the index 0 denotes the central 
bifurcating orbit and the index 1 the satellite orbit.
For the case $m=1$ there is no central periodic orbit
and the two periodic orbits which are involved in
the bifurcation are given the indices 1 and 2.
The relations (\ref{sec2i}) follow from the
normal forms and are valid in the vicinity of
a bifurcation. In farther distance from a bifurcation,
however, they are not valid any more, and then
the local approximation becomes inaccurate.
In order to obtain a formula which uniformly
interpolates over the region from the 
bifurcation up to regimes where the Gutzwiller
approximation is valid (without restrictions on
the semiclassical amplitudes) one has to include
higher order corrections to the normal forms in
(\ref{sec2h}). The resulting integrals can then
be reduced to simpler forms by appropriate 
coordinate transformations. These calculations
are carried out in appendices \ref{secm1},
\ref{secm2} and \ref{secm3}, and the results 
are discussed in the next section.

\section{Results for the uniform approximations}
\label{secres}

\subsection{The isochronuous bifurcation}

The bifurcation which is described by the normal
form in (\ref{sec2h}) for $m=1$ has the following
property: On one side of the bifurcation where
$\varepsilon$ and $a$ have opposite signs
there exist two periodic orbits, one stable and
one unstable.  We denote these orbits by $\xi_1$
and $\xi_2$.
On the other side of the bifurcation
where $\varepsilon$ and $a$ have the same sign
both orbits are complex, and we give the index 1
to that orbit for which the imaginary part of
the action is positive. Due to the shape of
the function $\hat{S}(q',p,E) - q' p$ this bifurcation
is called tangent bifurcation or saddle-node
bifurcation.

The uniform approximation for the semiclassical
contributions of the two orbits is derived in
appendix \ref{secm1}. Its form is different
on the two sides of the bifurcation. On the
side where the orbits are real it is given by
\begin{eqnarray} \label{sec3a}
d_\xi(E) &=& \frac{1}{\pi\hbar} 
\left|\frac{2\pi\Delta S}{3\hbar}\right|^{1/2}
\left\{ \frac{A_1 + A_2}{2} \, \cos \left( \frac{\bar S}{\hbar}
- \frac{\pi}{2} \bar{\nu} \right)
\left( J_{-1/3} \left(\frac{|\Delta S|}{\hbar}\right)
+      J_{ 1/3} \left(\frac{|\Delta S|}{\hbar}\right) \right)
\right. \\ \nonumber && \left.
- \mbox{sign}(\Delta S) \, \frac{A_1 - A_2}{2} \, 
\cos \left( \frac{\bar S}{\hbar}
- \frac{\pi}{2} (\bar{\nu} - 1) \right)
\left( J_{-2/3} \left(\frac{|\Delta S|}{\hbar}\right)
-      J_{ 2/3} \left(\frac{|\Delta S|}{\hbar}\right) \right) 
\right\} \; ,
\end{eqnarray}
which is invariant under exchange of the two indices.
The quantities which appear in (\ref{sec3a}) are
the mean action $\bar{S} = (S_1 + S_2)/2$ , the
action difference $\Delta S = (S_1 - S_2)/2$ and
the mean Maslov index $\bar{\nu} = (\nu_1 + \nu_2)/2
= \nu + \sigma/2$ of the orbits. Here $\nu$ is the 
index in (\ref{sec2f}) and $\sigma$ is the sign factor in
the normal form (\ref{sec2h}). Furthermore, $A_i$ denotes
here and in the following the semiclassical amplitude of
an orbit
\begin{equation} \label{sec3b}
A_i = \frac{T_i}{r_i \sqrt{\eta(\mbox{Tr \,} M_i - 2)}} \, ,
\end{equation} 
where $\eta = \mbox{sign\,}(\mbox{\bf Re\,}(\mbox{Tr\,} M_i - 2))$.
Properties of the two orbits when they are real
are listed in table 1, 
and an expansion of the classical properties of the
orbits in terms of the coefficients in the (extended)
normal form are given in appendix~\ref{secm1}. 
\begin{table}[htbp]
\begin{center}
\begin{tabular}{|c|ll|ll|}
\hline
{\large $m=1$} & \multicolumn{2}{c|}{\large $\sigma > 0$} 
               & \multicolumn{2}{c|}{\large $\sigma < 0$ } 
\rule{0ex}{2.7ex} \\[0.3ex]
\hline
& $\xi_u$ unstable, & $\nu_u = \nu$ 
& $\xi_u$ unstable, & $\nu_u = \nu$ \rule{0ex}{2.7ex} \\
{\large $a \varepsilon < 0$}
& $\xi_s$ stable,   & $\nu_s = \nu+1$
& $\xi_s$ stable,   & $\nu_s = \nu-1$ \\[0.3ex]
& $S_s > S_u$ &  
& $S_u > S_s$ & \\[0.3ex]
\hline
\end{tabular}
\end{center}
\caption{Properties of real orbits that participate in a
generic isochronuous bifurcation ($m=1$).
The orbits are denoted by $\xi_u$ and $\xi_s$ where
the indices $u$ and $s$ denote the unstable and
stable orbit, respectively.}
\end{table}

On the other side of the bifurcation where the orbits
are complex the semiclassical contribution to the
level density is given by
\begin{eqnarray} \label{sec3c}
d_\xi(E) &=& \mbox{\bf Re \,} \left[ \frac{1}{\pi\hbar} 
\left|\frac{2 \Delta S}{\pi \hbar}\right|^{1/2}
\exp  \left( \frac{i}{\hbar} \bar{S}
- \frac{i \pi}{2} \nu - \frac{i \pi}{4} \sigma \right)
\right. \nonumber \\ && \left. \times
\left\{ \frac{A_1 + A_2}{2} \, K_{1/3} 
\left(\frac{|\Delta S|}{\hbar}\right)
+      \frac{A_1 - A_2}{2} \, K_{2/3} 
\left(\frac{|\Delta S|}{\hbar}\right)
\right\}  \right] \; ,
\end{eqnarray}
where now the actions and amplitudes are complex.
Again $\nu$ is the index in (\ref{sec2f}) and $\sigma$
is the sign factor in the normal form (\ref{sec2h}).
Both equations (\ref{sec3a}) and (\ref{sec3c}) can be
written in a combined form by expressing them in terms
of an Airy function.

In the limit $\varepsilon \rightarrow 0$ the leading order
semiclassical contribution of equations (\ref{sec3a}) and
(\ref{sec3c}) is given by
\begin{equation} \label{sec3ca}
d_\xi(E) = \frac{T_0 \Gamma(1/3)}{
\pi l \sqrt{6 \pi} \hbar^{7/6} a^{1/3}} 
\cos \left( \frac{S_0}{\hbar}
- \frac{\pi}{2} \nu - \frac{\pi}{2} \sigma \right) \; .
\end{equation}
Here $T_0$ and $S_0$ are, respectively, the period and action
of the orbits at the bifurcation,
$a$ is the coefficient in the normal form (\ref{sec2h}),
and $l$ is the repetition number of the bifurcating
orbits. The contribution (\ref{sec3ca}) is of order 
$\hbar^{-1/6}$ larger than the contribution of an
isolated period orbit. All appearing classical 
quantities in (\ref{sec3ca}) depend on the integer $l$.
In detail, $T_{0,l} = l T_{0,l=1}$, $S_{0,l} = l S_{0,l=1}$, 
$\nu_l = l \nu_{l=1}$ and $a_l = l^{5/2} a_{l=1}$.
The relation for the Maslov index follows from
the Maslov index of the unstable orbit in table~1
for multiple traversals.
The last relation is obtained by noting that all
coefficient in the normal (\ref{sec2h}) increase by a factor $l$
as $l$ is increased. However, $\sigma$ is restricted
to be $\pm 1$. This is achieved by a subsequent
(canonical) transformation $p \rightarrow p/\sqrt{l}$
and $q \rightarrow q \sqrt{l}$. It follows that the amplitude
of the contribution at the bifurcation decreases
like $l^{-5/6}$ with $l$.

\subsection{The period-doubling bifurcation}

The period-doubling bifurcation which is described
by the normal form in (\ref{sec2h}) for $m=2$ has
the following form: On one side of the bifurcation where
$\varepsilon$ and $a$ have the same sign there
is only one orbit which is called the central orbit.
This orbit changes its stability at the bifurcation
from stable to unstable or vice versa, and a new
orbit appears which is named the satellite orbit
since the two fixed points of the Poincar\'e map
which belong to this orbit lie symmetrically on
both sides of the fixed point of the central orbit.
This bifurcation is also called pitchfork bifurcation.

The uniform approximation for the contributions of 
these orbits to the level density is derived in
appendix \ref{secm2} and is given by
\begin{eqnarray} \label{sec3d}
d_\xi(E) &=& \mbox{\bf Re \,} \left[ \frac{1}{\pi\hbar}
\left|\frac{\pi \Delta S}{2 \hbar}\right|^{1/2}
\exp \left(\frac{i}{\hbar} \bar{S}
- \frac{i \pi}{2} \nu - \frac{i \pi}{4} \sigma 
\right)  \right. \\
&&\times \left\{\left(\frac{A_1}{2}+\frac{A_0}{\sqrt{2}}\right)
\left(\sigma_2 J_{1/4}\left(\frac{|\Delta S|}{\hbar}\right)
e^{i\sigma_1\pi/8}
+J_{-1/4}\left(\frac{|\Delta S|}{\hbar}\right)
e^{-i\sigma_1\pi/8}
\right)\right.
\nonumber\\
&& \left. \left.
+\left(\frac{A_1}{2}-\frac{A_0}{\sqrt 2}\right)\left(
J_{3/4}\left(\frac{|\Delta S|}{\hbar}\right)
e^{i\sigma_1 3\pi/8}
 +\sigma_2J_{-3/4}\left(\frac{|\Delta S|}{\hbar}\right)
e^{-i\sigma_1 3\pi/8}
\right)
\right\} \right] \;. \nonumber
\end{eqnarray}
This approximation is valid on both sides of the bifurcation.
Here $\bar{S} = (S_1 + S_0)/2$, $\Delta S = (S_1 - S_0)/2$,
$\nu$ is the index in (\ref{sec2f}) and $\sigma$ is the
sign factor in the normal form (\ref{sec2h}). The values
of $\nu$ and $\sigma$ can be determined from the properties
of the orbits which are listed in table 2. Furthermore
$\sigma_1 = \mbox{sign}(\Delta S)$ and $\sigma_2$ is a
sign factor which discriminates between both sides of 
the bifurcation. It is $1$ when both orbits are real
and $-1$ when only the central orbit is real. The orbit
$\xi_1$ contributes also to (\ref{sec3d}) when it is
complex, but its action and amplitude factor are always
real. The expression of the classical properties of
the two orbits in terms of the coefficients in the
(extended) normal form are given in appendix \ref{secm2}.
\begin{table}[htbp]
\begin{center}
\begin{tabular}{|c|lll|lll|}
\hline
{\large $m=2$} & \multicolumn{3}{c|}{\large $\sigma > 0$} 
               & \multicolumn{3}{c|}{\large $\sigma < 0$ } 
\rule{0ex}{2.7ex} \\[0.3ex]
\hline
& $\varepsilon > 0:$ & $\xi_0$ stable,   & $\nu_0 = \nu + 1$ 
& $\varepsilon > 0:$ & $\xi_0$ unstable, & $\nu_0 = \nu$ 
\rule{0ex}{2.7ex} \\[0.8ex]
{\large $a>0$}
& $\varepsilon < 0:$ & $\xi_0$ unstable, & $\nu_0 = \nu$
& $\varepsilon < 0:$ & $\xi_0$ stable,   & $\nu_0 = \nu-1$ \\
&                    & $\xi_1$ stable,   & $\nu_1 = \nu+1$
&                    & $\xi_1$ unstable, & $\nu_1 = \nu$ \\[0.15ex]
&                    & $S_1 > S_0$       &               
&                    & $S_1 > S_0$       &               \\[0.3ex]
\hline
& $\varepsilon < 0:$ & $\xi_0$ unstable, & $\nu_0 = \nu$ 
& $\varepsilon < 0:$ & $\xi_0$ stable,   & $\nu_0 = \nu-1$ 
\rule{0ex}{2.7ex} \\[0.8ex]
{\large $a<0$}
& $\varepsilon > 0:$ & $\xi_0$ stable,   & $\nu_0 = \nu+1$
& $\varepsilon > 0:$ & $\xi_0$ unstable, & $\nu_0 = \nu$ \\
&                    & $\xi_1$ unstable, & $\nu_1 = \nu$
&                    & $\xi_1$ stable,   & $\nu_1 = \nu-1$ \\[0.15ex]
&                    & $S_1 < S_0$       &               
&                    & $S_1 < S_0$       &               \\[0.3ex]
\hline
\end{tabular}
\end{center}
\caption{Properties of real orbits that participate in a
generic period-doubling bifurcation ($m=2$). The central
orbit is denoted by $\xi_0$ and the satellite orbit by
$\xi_1$.}
\end{table}

In the limit $\varepsilon \rightarrow 0$ the leading order
semiclassical contribution of equation (\ref{sec3d}) is given by
\begin{equation} \label{sec3da}
d_\xi(E) = \frac{T_0 \Gamma(1/4)}{
4 \pi l \sqrt{2 \pi} \hbar^{5/4} a^{1/4}} 
\cos \left( \frac{S_0}{\hbar}
- \frac{\pi}{2} \nu - \frac{\pi}{4} \sigma 
- \frac{\pi}{8} \sigma_1 \right) \; .
\end{equation}
Here $T_0$ and $S_0$ are, respectively, the period and action
of the orbits at the bifurcation,
$a$ is the coefficient in the normal form (\ref{sec2h}),
and $l$ is the repetition number of the orbit $\xi_1$.
The contribution (\ref{sec3da}) is of order 
$\hbar^{-1/4}$ larger than the contribution of an
isolated period orbit. The dependence of the classical 
quantities in (\ref{sec3da}) on the integer $l$
are given by: $T_{0,l} = l T_{0,l=1}$, $S_{0,l} = l S_{0,l=1}$, 
$\nu_l = l \nu_{l=1}$ and $a_l = l^3 a_{l=1}$.
The relation for the Maslov index follows from
the Maslov index of the unstable orbit in table~2
for multiple traversals, and the relation for the
coefficient $a$ from considerations analogous to those
for the isochronuous bifurcation. It follows that the amplitude
of the contribution at the bifurcation decreases
like $l^{-3/4}$ with $l$.

\subsection{The period-tripling bifurcation}

The period-tripling bifurcation is described by the
normal form in (\ref{sec2h}) for $m=3$. It involves
two orbits, the central orbit $\xi_0$ and
the satellite orbit $\xi_1$. Both orbits exist
before and after the bifurcation. As $\varepsilon$
goes through 0 both orbits approach each other,
they coincide at the bifurcation and then separate
again. For that reason this bifurcation has also
been named ``touch and go'' bifurcation \cite{SD96}.

The uniform approximation for this bifurcation is
derived in appendix \ref{secm3} and is given by
\begin{eqnarray} \label{sec3e}
d_\xi(E) &=& \frac{1}{\pi \hbar} \, \mbox{\bf Re} \,
\sqrt{\frac{2 \pi |\Delta S|}{\hbar}} \,
\exp \left\{ \frac{i}{\hbar} \bar{S} 
- \frac{i \pi}{2} \nu \right\} \, 
\nonumber \\ &&
\times \left\{ 
\left( \frac{A_0}{2} + \frac{A_1}{2 \sqrt{3}} \right)
\left[ \J_{-1/6} 
\left( \frac{|\Delta S|}{\hbar} \right)
+ i \sigma \J_{1/6} 
\left( \frac{|\Delta S|}{\hbar} \right)
\right] \right.
\nonumber \\ &&
\left. -
\left( \frac{A_0}{2} - \frac{A_1}{2 \sqrt{3}} \right)
\left[ \J_{-5/6} 
\left( \frac{|\Delta S|}{\hbar} \right)
+ i \sigma \J_{5/6} 
\left( \frac{|\Delta S|}{\hbar} \right)
\right] \right\} \; .
\end{eqnarray}
Here $\bar{S} = (S_1 + S_0)/2$, $\Delta S = (S_1 - S_0)/2$,
$\nu=\nu_1$ is the Maslov index of the satellite orbit.
Furthermore $\sigma = \mbox{sign}(\Delta S)$. Properties
of the two orbits are listed in table 3, and the
expression of the classical properties of
the two orbits in terms of the coefficients in the
(extended) normal form are given in appendix \ref{secm3}.
\begin{table}[htbp]
\begin{center}
\begin{tabular}{|c|ll|ll|}
\hline
{\large $m=3$} & \multicolumn{2}{c|}{\large $\varepsilon > 0$} 
               & \multicolumn{2}{c|}{\large $\varepsilon < 0$ } 
\rule{0ex}{2.7ex} \\[0.3ex]
\hline
& $\xi_0$ stable,   & $\nu_0 = \nu+1$
& $\xi_0$ stable,   & $\nu_0 = \nu-1$ \\
& $\xi_1$ unstable, & $\nu_1 = \nu$
& $\xi_1$ unstable, & $\nu_1 = \nu$ \\[0.15ex]
& $S_1 < S_0$       &               
& $S_1 > S_0$       &               \\[0.3ex]
\hline
\end{tabular}
\end{center}
\caption{Properties of orbits that participate in a
generic period-tripling bifurcation ($m=3$). The central
orbit is denoted by $\xi_0$ and the satellite orbit by
$\xi_1$.}
\end{table}

In the limit $\varepsilon \rightarrow 0$ the leading order
semiclassical contribution of equation (\ref{sec3e}) is given by
\begin{equation} \label{sec3ea}
d_\xi(E) = \frac{T_0 \Gamma(1/6)}{
9 l \pi^{3/2} \hbar^{4/3} a^{2/3}} 
\cos \left( \frac{S_0}{\hbar}
- \frac{\pi}{2} \nu \right) \; .
\end{equation}
Here $T_0$ and $S_0$ are, respectively, the period and action
of the orbits at the bifurcation,
$a$ is the coefficient in the normal form (\ref{sec2h}),
and $l$ is the repetition number of the orbit $\xi_1$.
The contribution (\ref{sec3ea}) is of order 
$\hbar^{-1/3}$ larger than the contribution of an
isolated period orbit. The dependence of the classical
quantities in (\ref{sec3ea}) on the integer $l$ are
given by: $T_{0,l} = l T_{0,l=1}$, $S_{0,l} = l S_{0,l=1}$, 
$\nu_l = l \nu_{l=1}$ and $a_l = l a_{l=1}$.
The relation for the Maslov index follows from
the Maslov index of the unstable orbit in table~3
for multiple traversals. It follows that the amplitude
of the contribution at the bifurcation decreases
like $l^{-2/3}$ with $l$. 

\section{Uniform approximations for maps}

Though we concentrated on autonomous systems in the preceding
paragraphs, the uniform contributions given there can equally
be applied, with minor modifications, to quantum maps. These
maps are described
by a unitary time-evolution operator $F$, the so-called Floquet
operator; the dynamics of the system is generated by repeated
applications of the operator on a state in Hilbert space. $F$ has
eigenstates and unimodular eigenvalues $e^{-i\varphi_i}$, where
the phases $\varphi_i$ are called {\em quasi energies} 
and the states {\em quasi-stationary states} since many quantum
maps originate from a stroboscopic description of periodically driven
systems. For such systems with Hamiltonian $H(t+T)=H(t)$, the unitary
time-evolution operator after period $T$ is chosen as $F=U(T)$,
implying $U(nT)=F^n$.

Restricting oneself onto a finite-dimensional subspace of Hilbert
space with dimension $N$ the quasi-energy spectrum can be
obtained by solving the
secular equation 
\[
P(\lambda=e^{-i\varphi}):=
\det(F - \lambda I)
=\sum_{n=0}^N a_n\lambda^{N-n}=0
\]
for the $N\times N$-matrix $F$. Here the set of traces  $\mbox{Tr}\,F^n$
comes into play:
their knowledge for $n$ up to $N/2$ allows to construct the first
half of the coefficients $a_n$ via the so-called Newton formulae;
the other half follows from the unitarity of $F$ which
entails self-inversiveness of the secular polynomial \cite{BBL92}.

The semiclassical starting point for the calculation of
$\mbox{Tr}\,F^n$ is an expression which is nearly identical
in its appearance
with equation (\ref{sec2f}) for autonomous systems,
\begin{equation}\label{sec5a}
\mbox{Tr}\,F^n=\int\!\!\int\frac{dp\,dq}{2\pi\hbar}
\left| \frac{\partial^2 \hat{S}}{\partial p \partial q'}
\right|^{\frac{1}{2}}
\exp\left\{\frac i \hbar
(\hat{S}(q^\prime,p;nT)-q^\prime p)
-i\frac \pi 2 \nu\right\}.
\end{equation}
The major difference to the expression for
autonomous systems is that one does not switch to energy 
representation by a Fourier transform with respect to time
$t$. In translating the contributions which were derived for
autonomous systems into the appropriate contributions to $\mbox{Tr}\,F^n$
attention has to be paid to the following five difference:
(i) The orbits which contribute are those with a fixed period $n$, not those
with a given energy $E$;  
(ii) the primitive periods have
to be expressed in units of $T$ and thus are integer valued;
(iii) the action is not the reduced energy dependent one,
but depends on time (that is, on the number $n$);
(iv) instead of taking twice the real part, the full complex contribution
has to be taken;
(v) the results further differ by a factor $2\pi\hbar$.
Since these are mainly formal differences
the morphology of the contributions remains unaltered.

In the limit $\hbar\to 0$
each orbit of primitive period $n_0$ 
contributes individually according to
\begin{equation}\label{sec5b}
\mbox{Tr}\,F^n=\sum_{\rm orbits} \frac {n_0}{|2-\mbox{Tr}\,M|^{1/2}}
\exp\left\{\frac i\hbar S-i\frac \pi 2 \nu\right\}\;.
\end{equation}
In the following we denote the contributions to $\mbox{Tr}\,F^n$
by $C^{(n)}$ and use $A_i= n_{0,i}(\eta(\mbox{Tr}\,M_i-2))^{-1/2}$
to abbreviate the stability amplitudes with
$\eta = \mbox{sign\,}(\mbox{\bf Re\,}(\mbox{Tr\,} M_i - 2))$.
$\bar S=(S_1+S_i)/2$ is the mean action and $\Delta S=(S_1-S_i)/2$ 
the action difference where $i$ is 0 or 2 depending on the considered
bifurcation, and $\nu$ is the index in (\ref{sec5a}).

For $m=1$ the collective contribution is
\begin{eqnarray} \label{sec5c}
C^{(n)} &=&
\left|\frac{2\pi\Delta S}{3\hbar}\right|^{1/2}
\left\{ \frac{A_1 + A_2}{2} \, \exp \left( i\frac{\bar S}{\hbar}
- i\frac{\pi}{2} \bar{\nu} \right)
\left( J_{-1/3} \left(\frac{|\Delta S|}{\hbar}\right)
+      J_{ 1/3} \left(\frac{|\Delta S|}{\hbar}\right) \right)
\right. \\ \nonumber && \left.
- \mbox{sign}(\Delta S) \, \frac{A_1 - A_2}{2} \, 
\exp \left( i\frac{\bar S}{\hbar}
- i\frac{\pi}{2} (\bar{\nu} - 1) \right)
\left( J_{-2/3} \left(\frac{|\Delta S|}{\hbar}\right)
-      J_{ 2/3} \left(\frac{|\Delta S|}{\hbar}\right) \right) 
\right\} \;
\end{eqnarray}
when both orbits are real.
When they are complex one obtains
\begin{eqnarray} \label{sec5d}
C^{(n)} &=&
\left|\frac{2 \Delta S}{\pi \hbar}\right|^{1/2}
\exp  \left( \frac{i}{\hbar} \bar{S}
- \frac{i \pi}{2} \nu - \frac{i \pi}{4} \sigma \right)
\nonumber \\ && \times
\left\{ \frac{A_1 + A_2}{2} \, K_{1/3} 
\left(\frac{|\Delta S|}{\hbar}\right)
+      \frac{A_1 - A_2}{2} \, K_{2/3} 
\left(\frac{|\Delta S|}{\hbar}\right)
\right\}  \; .
\end{eqnarray}
The amplitudes $A_i$ are now complex quantities. 
$\sigma$ is the sign factor in the normal form (\ref{sec2h}).
A pair of orbits involved in a period doubling bifurcation ($m=2$)
gives the contribution
\begin{eqnarray} \label{sec5e}
C^{(n)} &=&
\left|\frac{\pi \Delta S}{2 \hbar}\right|^{1/2}
\exp \left(\frac{i}{\hbar} \bar{S}
- \frac{i \pi}{2} \nu - \frac{i \pi}{4} \sigma 
\right)  \\
&&\times \left\{\left(\frac{A_1}{2}+\frac{A_0}{\sqrt{2}}\right)
\left(\sigma_2 J_{1/4}\left(\frac{|\Delta S|}{\hbar}\right)
e^{i\sigma_1\pi/8}
+J_{-1/4}\left(\frac{|\Delta S|}{\hbar}\right)
e^{-i\sigma_1\pi/8}
\right)\right.
\nonumber\\
&& \left.
+\left(\frac{A_1}{2}-\frac{A_0}{\sqrt 2}\right)\left(
J_{3/4}\left(\frac{|\Delta S|}{\hbar}\right)
e^{i\sigma_1 3\pi/8}
 +\sigma_2J_{-3/4}\left(\frac{|\Delta S|}{\hbar}\right)
e^{-i\sigma_1 3\pi/8}
\right)
\right\} \nonumber
\end{eqnarray}
with $\sigma_1=\mbox{sign}\,(\Delta S)$ and $\sigma_2=1$ when the
satellite is real, $\sigma_2=-1$ otherwise.
For $m=3$ the result reads
\begin{eqnarray} \label{sec5f}
C^{(n)} &=&
\sqrt{\frac{2 \pi |\Delta S|}{\hbar}} \,
\exp \left\{ \frac{i}{\hbar} \bar{S} 
- \frac{i \pi}{2} \nu \right\} \, 
\nonumber \\ &&
\times \left\{ 
\left( \frac{A_0}{2} + \frac{A_1}{2 \sqrt{3}} \right)
\left[ \J_{-1/6} 
\left( \frac{|\Delta S|}{\hbar} \right)
+ i \sigma \J_{1/6} 
\left( \frac{|\Delta S|}{\hbar} \right)
\right] \right.
\nonumber \\ &&
\left. -
\left( \frac{A_0}{2} - \frac{A_1}{2 \sqrt{3}} \right)
\left[ \J_{-5/6} 
\left( \frac{|\Delta S|}{\hbar} \right)
+ i \sigma \J_{5/6} 
\left( \frac{|\Delta S|}{\hbar} \right)
\right] \right\} 
\end{eqnarray}
with  $\sigma=\mbox{sign}\,(\Delta S)$.

These expressions give also the
contribution of repetitions of the bifurcating
orbits if one substitutes the corresponding classical quantities.
For the $l$-th repetition, the stability angle $\omega$ of a
stable orbit in $\mbox{Tr}\,M=2\cos\omega$ and the instability
exponent $u$ of an unstable orbit in $\mbox{Tr}\,M=\pm 2\cosh u$
increase linearly,
$\omega_l=l\omega_{l=1}$ and $u_l = l u_{l=1}$, 
as does the action $S_l=l S_{l=1}$ and the Maslov index
of the unstable orbits $\nu_l^{(u)}=l\nu_{l=1}^{(u)}$.
The Maslov index of the stable orbits is 
$\nu_l^{(s)} = \nu_l^{(u)} + \mbox{sign}(S_l^{(s)}-S_l^{(u)})$.

\section{Numerical results}

In this section we want to test the quality
of the uniform collective contributions
that we derived for the various types of bifurcations on
a certain quantum map, a periodically kicked top \cite{HKS87,KHE93,BGHS96}. 
It will turn out that the uniform approximations indeed
are accurate both close to the bifurcation as well as
in far distance; the local
approximation is only valid close to the bifurcation
while far away 
the orbits can be treated as being isolated 
via the stationary-phase approximation.

Tops are dynamical systems that involve the
angular momentum operators
$J_x$, $J_y$, $J_z$, satisfying the usual commutation relations
$[J_k,J_l]=i\epsilon_{klm}J_m$, where $\hbar$ is set to unity. 
The evolution of the system is such that
the total angular momentum $J_x^2+J_y^2+J_z^2=j(j+1)$ is conserved.
This introduces the
well known good quantum number $j$ which 
fixes the Hilbert space dimension $2j+1$.
$j+1/2$ further plays
the role of the inverse of Planck's constant; accordingly, the semiclassical
limit is reached by sending $j\to\infty$.

The specific top used here is described by the Floquet operator
\begin{equation}
F=\exp(-i\frac {k_z}{2j+1}J_z^2-ip_zJ_z)\exp(-ip_yJ_y)
\exp(-i\frac {k_x}{2j+1}J_x^2-ip_xJ_x)\;.
\end{equation}
The dynamics consists of rotations by angles $p_i$ and nonlinear
rotations (torsions) of strength $k_i$.
For the study of bifurcations we hold the $p_i$
at fixed values ($p_x=0.3$, $p_y=1.0$, $p_z=0.8$)
and take $k=k_z$ as our control parameter,
with $k_x=k/10$. The classical counterpart of the system
is integrable at $k=0$ and displays well developed chaos 
at $k=5$.

There is a very convenient
testing tool which enables one to examine each contribution
of a given cluster of periodic orbits
individually. It is obtained by considering the
function
\begin{equation}
T^{(n)}(S)=\frac 1 {j_{\rm max}-j_{\rm min}+1}  \sum_{j=j_{\rm min}}^{j_{\rm
max}}e^{-ijS}\mbox{Tr}\,F^n(j)\;.
\end{equation}
In its essence this
is a Fourier coefficient
of $\mbox{Tr}\,F^n$ with respect to $j$; finite limits have to be taken 
for practical reasons if one wants to evaluate the sum for the quantum system, 
they further give control over the desired degree of
rigor of the semiclassical limit.
We used $j_{\rm min}=1$ and $j_{\rm max}=64$ with a single exception for
$m=3$.

\begin{figure}[htbp]
\epsfxsize14cm
\epsfbox{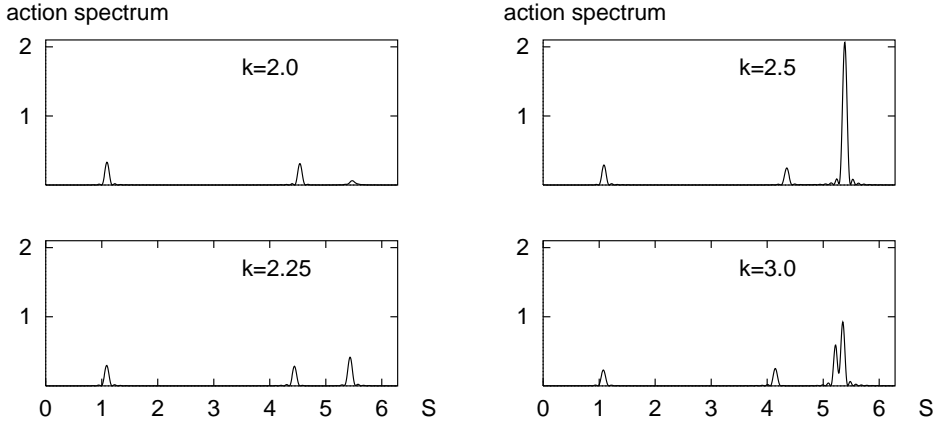}
\caption{Sequence of action spectra $|T^{(1)}(S)|^2$
as the control parameter crosses a tangent bifurcation ($m=1$) 
at $k=2.45$. A ghost peak shows up already before the bifurcation 
at $S=5.3$; beyond the bifurcation
the peak splits into two, each corresponding to a bifurcating orbit.
}
\label{fig1}
\end{figure}
\begin{figure}[htbp]
\epsfxsize4.5cm
\rotate[r]{\epsfbox{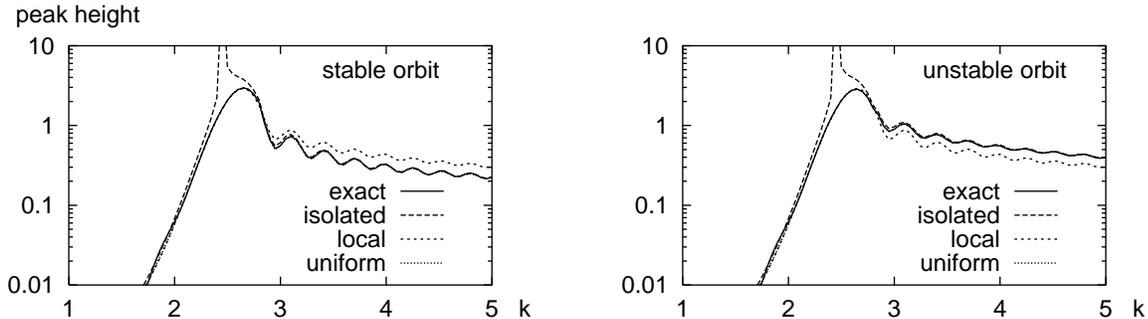}}
\caption{Height of the peaks in the action spectrum at the positions
of the two orbits that are engaged in the bifurcation of
Figure \protect\ref{fig1}. Both the uniform and the local bifurcation
work well close to and on the left of the bifurcation. 
The local approximation starts to fail as $k$ is increased, and
the sum of isolated contributions of Gutzwiller's type 
becomes valid.
Note that the uniform approximation remains accurate and
can hardly be
distinguished from the exact result in this plot.
}
\label{fig2}
\end{figure}
{}From the asymptotic behaviour (\ref{sec5b}) of all uniform
approximations
it is clear that one expects peaks in $|T^{(n)}(S)|^2$ at values of the
argument which correspond to actions of periodic orbits.
The function $|T^{(n)}(S)|^2$ is called the action spectrum for that reason.
Figure \ref{fig1} confirms that peaks indeed show up.
Plotted is a sequence of action spectra for $n=1$ as the control
parameter $k$ is varied across a tangent bifurcation at
$k=2.45$. Already before the pair of orbits comes into existence
a peak is visible at $S=5.3$. This peak  arises from the complex
predecessors of the bifurcating orbits that were already observed in 
\cite{KHD93}. The other peaks pertain to different orbits.
Slightly beyond the bifurcation the new-born orbits have nearly 
identical actions and give rise to a single peak that would 
be resolved if one would go to much higher $j_{\max}$.
Increasing the control parameter further the peak splits into two 
peaks which are located at the
now well separated values of the orbits' actions.

The quality of our results can now be tested by calculating the
height of the peaks $|T^{(n)}(S_{\rm cl})|^2$ both quantum-mechanically
exact as well as on grounds of the various semiclassical approximations.
Figure \ref{fig2} depicts the height of the aforementioned peaks
as
the control parameter is varied across the tangent bifurcation.
In the vicinity of the bifurcation both the local as well as
the uniform approximation are accurate to a degree
that makes it difficult to resolve the error
at all. 
The uniform approximation remains also valid as one moves away
from the bifurcation.
There the sum of individual contributions of isolated orbits is seen
to gain validity.  

\begin{figure}[htbp]
\epsfxsize4.5cm
\rotate[r]{\epsfbox{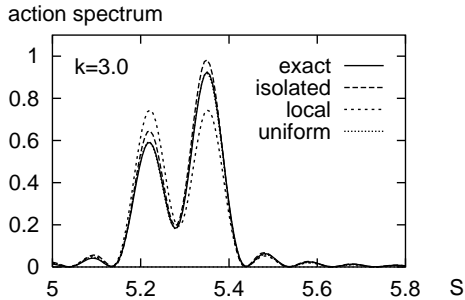}}
\caption{The exact peak in the action spectrum 
of Figure \protect\ref{fig1} at $k=3.0$ is compared
to semiclassically evaluated ones.
The uniform approximation works well. The
local approximation predicts peaks of equal height.
Treating both orbits as isolated gives peaks that are slightly too high.
}
\label{fig3}
\end{figure}
Figure \ref{fig3} illustrates the superiority of the uniform approximation
over the local one away from the bifurcation. To that end,
the exact peak itself is compared to the approximated ones.
The local approximation  assumes  that the stabilities of both orbits are
equal and gives two peaks of same height. With the sum of two isolated
terms of Gutzwiller's type
both peak heights are slightly overestimated.
The uniform approximation almost coincides with the exact result.

\begin{figure}[htbp]
\epsfxsize4.5cm
\rotate[r]{\epsfbox{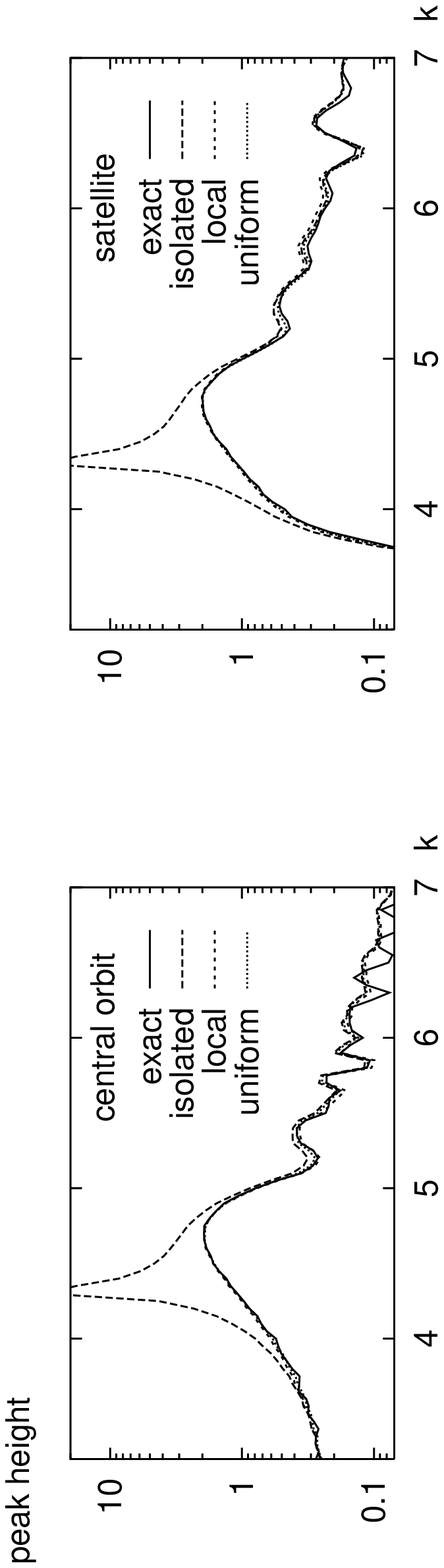}}
\caption{Peak height as in Figure \protect\ref{fig2} for orbits involved
in a period doubling bifurcation ($m=2$)
at $k=4.3$. The local and uniform approximations are again
accurate before and close to the bifurcation. This time they remain
so until overlapping peaks from other orbits give rise to deviations.
One such orbit has already been included, others not, as is explained in
the text.
The sum of individual terms that treats all orbits as isolated
fails close to the bifurcation and regains validity far away.
}
\label{fig4}
\end{figure}
\begin{figure}[htbp]
\epsfxsize4.5cm
\rotate[r]{\epsfbox{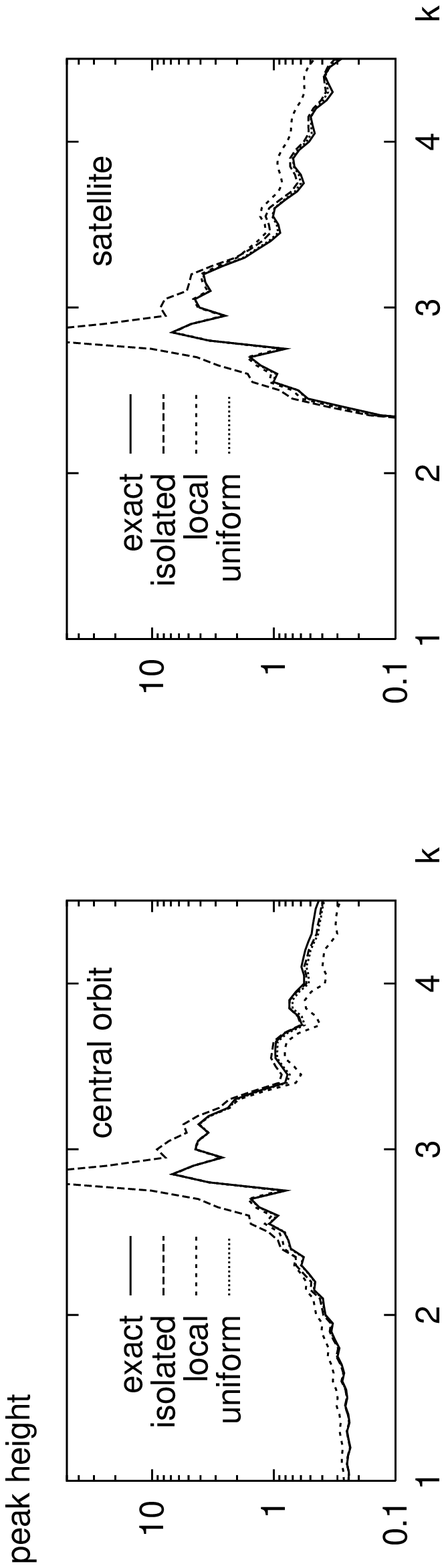}}
\caption{Two pairs of orbits involved in period doublings
at $k=2.78$ and $k=2.84$ 
give rise to overlapping peaks in the action spectrum; the height at
the actions of two bifurcating orbits is plotted here. 
The accuracy of the uniform approximation in the whole parameter range,
the failure of the isolated treatment close the bifurcation, and
the breakdown of the local approximation at a distance to the bifurcation
is clearly visible.
}
\label{fig5}
\end{figure}
In Figure \ref{fig4}, the
peak heights of orbits participating in a period doubling at $k=4.3$
are plotted. In this case it is hardly possible to discriminate
between the local and the uniform approximations. Before
the amplitudes start to differ significantly, erratic deviations
to the exact result are encountered which stem from overlapping peaks of
other orbits.
One such orbit has already been included here with its 
isolated contribution for $k>5.8$ and cures part of the problem, but
other orbits become relevant from $k\approx 6.2$. Instead of further
improving on this we settle for the information that
the approximations work well close to the bifurcation and turn to
other period doublings.
Figure \ref{fig5} actually arises from a similar
situation of two overlapping peaks; both peaks, however, arise from
two pairs of orbits which undergo period doublings at $k=2.78$ and
$k=2.84$, respectively. That both pairs have almost the same action is a
pure coincidence.
Plotted is now the collective contribution
of the total number of four orbits in various semiclassical
approximations and the exact result. The situation is
qualitatively the same as for the tangent bifurcation: both the local
and the uniform approximation are excellent close to the bifurcations;
moving away the uniform approximation remains valid and the sum of isolated
orbits starts to work well while the local approximation breaks down.

\begin{figure}[htbp]
\epsfxsize4.5cm
\rotate[r]{\epsfbox{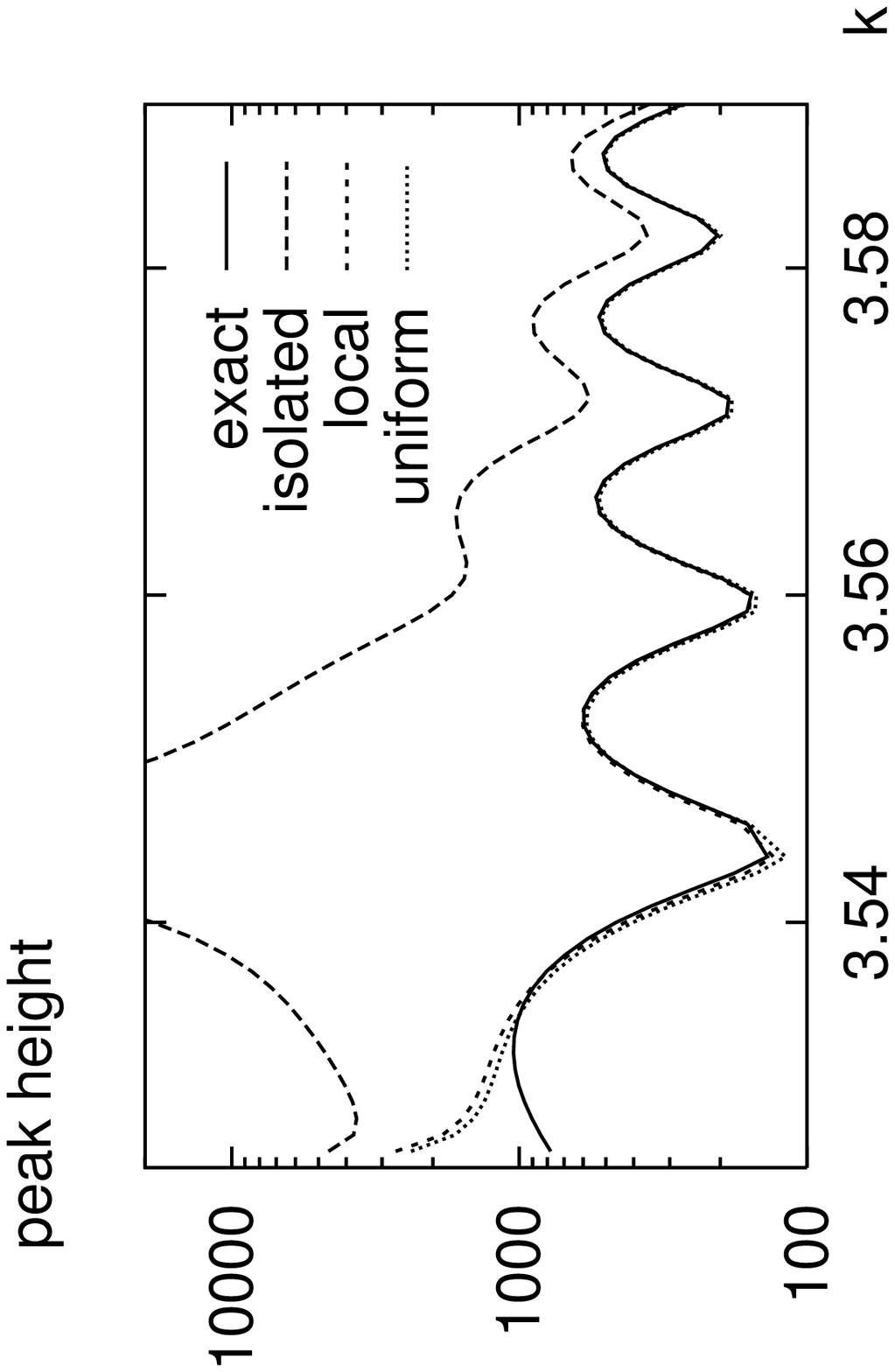}}
\caption{Height of peaks in the action spectrum  
that arise from orbits involved in a period tripling bifurcation ($m=3$)
at $k=3.545$. The satellite is engaged in a tangent bifurcation
with another satellite
at $k=3.525$; this requires to increase $j_{\rm min}=2^{13}+1$
as discussed in the text. The additional satellite is only
felt slightly at the bifurcation and the accuracy is high for the
local and uniform approximation.
}
\label{fig6}
\end{figure}

\begin{figure}[htbp]
\epsfxsize4.5cm
\rotate[r]{\epsfbox{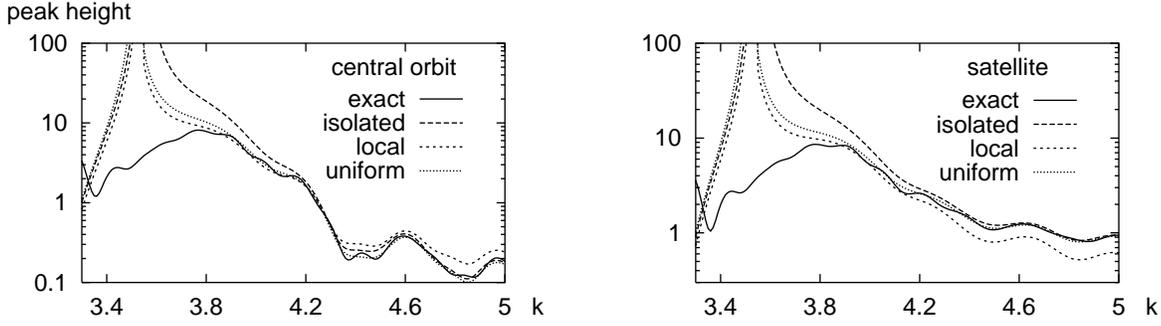}}
\caption{The same period tripling is investigated as in Figure
\protect\ref{fig6}. For the chosen $j_{\rm min}=1$ an
additional satellite
overshadows the result close to the tripling.
As one goes away from the bifurcation to higher $k$ the uniform and
isolated approximations work fine and the local approximation
is inaccurate.}
\label{fig7}
\end{figure}
For the period tripling we hit on difficulties from the close
neighborhood of a tangent bifurcation of the satellite. This tangent
bifurcation was found to be very close to the tripling for all
investigated cases. The problem can be overcome if
$j_{\min}$ is chosen large enough in
order to ensure that the further satellite which is involved into 
the tangent bifurcation is well separated, the separation being measured
by its difference in action in units of Planck's constant, 
$(j+1/2)\Delta S$.
Then our analytical
expression
(\ref{sec5f}), which does not describe the tangent bifurcation,
can be tested without being overshadowed by the existence of the
additional satellite.
Since the relevant action difference
is tiny we raised $j_{\rm min}$ up
to $2^{13}+1$, with $j_{\rm max}=2^{13}+64$.
The result is presented in Figure \ref{fig6}. For the lower values
of $k$ the tangent bifurcation at $k=3.525$
is still felt strongly and all
semiclassical approximations go wrong. At the period tripling
($k=3.545$) 
and beyond it the uniform and the local
approximation work well again.
The breakdown of the local approximation is observed if one
goes to a different scale of the control parameter, as shown
in Figure \ref{fig7}.

In conclusion, for all studied cases the uniform approximation was found
to give excellent results (as long as the
results were not overshadowed by the existence of further orbits).
The local approximation is of the same quality close to the
bifurcation. 
It breaks down when the relations (\ref{sec2i}) between the monodromy matrices
become invalid as one increasingly detunes the control
parameter.
Along the same line, the action difference of the orbits in units of
Planck's constant is large in that region of parameter space,
and the use of the sum over isolated orbits (\ref{sec5b}) makes sense there.

\section{Conclusions}

In this paper we examined semiclassical contributions of
three different bifurcations in two-dimensional conservative
systems and one-dimensional area-preserving maps. These
bifurcations are the generic period-$m$-tupling bifurcations with
$m=1$, 2 and 3. We extended the local approximation
of Ozorio de Almeida and Hannay for these cases
and derived uniform approximations which interpolate over the
regime from the bifurcation up to regions where the orbits
can be considered isolated. The approximations were tested
numerically on the example of a kicked top and were found
to be in excellent agreement with exact quantum calculations.
The local approximation was found to be good near a bifurcation,
but its quality decreased in farther distance from a
bifurcation. In contrast, the Gutzwiller approximation is
good sufficiently far away from a bifurcation, but it
diverges at a bifurcation.

We derived the classical properties of periodic orbits
that follow from normal forms with higher order corrections.
We further examined the semiclassical contributions of the
orbits directly at the bifurcation ($\varepsilon = 0$).
It was found that with increasing $m$ bifurcations
contribute more strongly to the level density.
In more detail, the semiclassical amplitude of 
the orbits at the bifurcation is of order
${\cal O}(\hbar^{-7/6})$ for $m=1$,
${\cal O}(\hbar^{-5/4})$ for $m=2$, and
${\cal O}(\hbar^{-4/3})$ for $m=3$. Furthermore,
for larger $m$ the semiclassical amplitude decreases
more slowly for higher repetitions of the bifurcating
periodic orbits. For the $l$-th multiple of the bifurcating
orbits the semiclassical amplitude decreases by a factor
$l^{-5/6}$ for $m=1$, $l^{-3/4}$ for $m=2$, and
$l^{-2/3}$ for $m=3$. For the case $m \geq 5$ it follows
from \cite{Sie96} that the semiclassical amplitude at
the bifurcation is of order ${\cal O}(\hbar^{-3/2})$
and it decreases by a factor $l^{-1/2}$ with increasing $l$.
Thus bifurcations with larger $m$ have
a stronger influence on the level density. 
 
All uniform approximations in this paper are valid as long
as the participating periodic orbits do not bifurcate further, 
i.\,e.\ as long as the bifurcations can be considered isolated. 
It can happen that a periodic orbit undergoes several subsequent
bifurcations. This has been observed in the numerical
examples where a tangent bifurcation occurred very close to
a period-tripling bifurcation. Similar situations for 
period-doubling and period-quadrupling bifurcations have
been observed in \cite{SSD95,SD96}. In these cases the
uniform approximation has to be modified by fully integrating
over an extended normal form in which higher order
terms are included that describe also subsequent
bifurcations. For the mentioned period-tripling bifurcation
this is done in \cite{HS97}.

An extension of the present results concerns
the case $m=4$, i.\,e.\ the period-quadrupling
bifurcation. Since the case $m>4$ was treated in
\cite{Sie96} this is the only remaining generic 
bifurcation. The treatment of the generic
period-quadrupling bifurcation is more complicated
than the other cases since it involves three
periodic orbits whose action differences all
scale with the same leading power in $\varepsilon$
(for small $\varepsilon$).
As a consequence one has to add two correction
terms to the local approximation in order to
obtain a formula which has the correct Gutzwiller
limit. Furthermore, the diffraction catastrophe
integral for this case can in general not be
expressed in terms of simple functions. A
treatment of this bifurcation is in preparation
\cite{SS97}.

A further extension concerns systems with symmetries.
In these systems there can be further kinds of bifurcations 
which are characteristic for the considered symmetry. 
Normal forms for systems with symmetries have been
derived in \cite{Rim82,AMBD87}. For some cases the
uniform approximations for bifurcations in systems
with symmetries can be obtained from a slight 
modification of the formulas for generic bifurcations.
For example, in systems with time-reversal and
reflection symmetries there can be isochronuous
pitchfork bifurcations. These bifurcations are described
by formula (\ref{sec3d}) with the replacement of $A_1$
by $2 A_1$. Another example where the generic formula
for $m \geq 5$ is applied to symmetric period-$n$-tupling
bifurcations with $n=m/2$ is discussed in \cite{Sie96b}.

\bigskip \bigskip \noindent {\large \bf Acknowledgments} 
\vspace{2.5mm}

M.\ Sieber wishes to acknowledge financial support
by the Deutsche Forschungsgemeinschaft under contract 
No.\ DFG-Ste 241/7-1 and /6-1. 
H.\ Schomerus gratefully acknowledges support by the
Sonderforschungsbereich `Unordnung und gro{\ss}e Fluktuationen' of the
Deutsche Forschungsgemeinschaft.

\appendix

\section{Uniform approximation for the isochronuous bifurcation} 
\label{secm1}

The starting point for the derivation of the uniform approximation
is the expansion of the generating function in the vicinity of
the bifurcation
\begin{equation} \label{m1a}
\hat{S}(q',p,E)= S_0 + q' p - \varepsilon q' - a q'^3 - b q'^4
- \frac{\sigma}{2} p^2 \, .
\end{equation}
Here we went one order higher in $q'$ than in the normal form
expansion, and in the following we will treat the extra term
as a perturbation to obtain the uniform approximation.
By rescaling $q'$ and $p$ we can enforce $\sigma=\pm 1$.
The fixed points of the map generated by $\hat{S}$
are given by $\hat{S}_{q'}=p$, $\hat{S}_p=q'$, and up to second order in 
$\sqrt{|\varepsilon|}$ we get for the two fixed points engaged
in the bifurcation
\begin{eqnarray} \label{m1b}
p_{1,2} &=& 0 \nonumber \\
q'_{1,2} &=& \pm \sqrt{-\frac{\varepsilon}{3a}} +
\frac{2b\varepsilon}{9a^2} +{\cal O}(|\varepsilon|^{3/2}) \; ,
\end{eqnarray}
while we will not consider the additional
fixed point which now formally arises as a third
solution of the equations.
The value of the actions $S=\hat{S}-q'p$ at the fixed points is
\begin{equation} \label{m1c}
S_{1,2}= S_0  \mp \frac{2\varepsilon}{3} \sqrt{-\frac{\varepsilon}{3a}}
- \frac{b\varepsilon^2}{9a^2} +{\cal O}(|\varepsilon|^{5/2}) \; ,
\end{equation}
where $S_0$ is the value of the generating function at the origin.
If $\varepsilon$ has the same sign as $a$ then the orbits are
complex and there is an ambiguity for choosing the sign of the
square root in Eqs.\ (\ref{m1b}) and (\ref{m1c}). We then choose
to give the index 1 to that orbit for which the imaginary part of
the action is positive. This formally corresponds to choosing
$\sqrt{-\varepsilon/(3a)} = - i \mbox{sign}(\varepsilon) 
\sqrt{|\varepsilon/(3a)|}$.

The periods of the orbits are given by 
\begin{equation} \label{m1d}
T_{1,2} = T_0 \mp \varepsilon_E \sqrt{\frac{-\varepsilon}{3a}}
+ {\cal O}(\varepsilon) \; ,
\end{equation}
where $T_0 = \partial \hat{S}/ \partial E$, evaluated at the origin,
and corresponds to the mean period. 
The traces of the monodromy matrices follow from the 
relation 
\begin{equation} \label{m1e}
\mbox{Tr} M = \left( 
\frac{\partial^2 \hat{S}}{\partial p \, \partial q'} 
\right)^{-1} \, \left( 1 + 
\frac{\partial^2 \hat{S}}{\partial p \, \partial q'} 
\frac{\partial^2 \hat{S}}{\partial p \, \partial q'} -
\frac{\partial^2 \hat{S}}{\partial p^2} 
\frac{\partial^2 \hat{S}}{\partial q'^2}
\right) \; ,
\end{equation}
which is evaluated at the stationary points and leads to
\begin{equation} \label{m1f}
\mbox{Tr\,} M_{1,2} = 2 \mp \sigma 6 a \sqrt{-\frac{\varepsilon}{3a}}
+ \frac{8 \sigma b}{3 a } \varepsilon 
+ {\cal O}(|\varepsilon|^{3/2}) \; .
\end{equation}
Furthermore the semiclassical amplitudes are given by
\begin{equation} \label{m1g}
A_{1,2} = \frac{T_{1,2}}{l \sqrt{\eta_{1,2}(\mbox{Tr\,} M_{1,2}-2)}}
=\frac{1}{l \left| 12 a \varepsilon \right|^{1/4}}
\left(T_0 \mp \left( \frac{2b}{3a} T_0 + \varepsilon_E \right)
\sqrt{-\frac{\varepsilon}{3a}}
+ {\cal O}(\varepsilon) \right) \; ,
\end{equation}
where $\eta_i = \mbox{sign\,}(\mbox{\bf Re\,}(\mbox{Tr\,} M_i - 2))$,
and $l=r_1=r_2$ is the repetition number of the orbits.

In the following we will use the definitions
\begin{equation} \label{m1h}
\bar{S}  = \frac{S_1 + S_2}{2} \, , \; \; \; \; 
\Delta S = \frac{S_1 - S_2}{2} \, , \; \; \; \; 
\bar{A}  = \frac{A_1 + A_2}{2} \, , \; \; \; \; 
\Delta A = \frac{A_1 - A_2}{2} \, , \; \; \; \; 
\bar{\nu}  = \frac{\nu_1 + \nu_2}{2} \, ,
\end{equation}
where $\nu_1$ and $\nu_2$ are the Maslov indices of the
orbits when they are real. 

We continue now with the evaluation of the integral in (\ref{sec2f}).
The main contribution to the integral over $q'$ comes from the
region near the stationary points. For that reason we consider
$q'$ in the following as a quantity of order 
${\cal O}(|\varepsilon|^{1/2})$. Then the exponent of the
integral can be simplified by substituting
\begin{equation} \label{m1i}
q' = \frac{\varepsilon b}{9a^2} + x - x^2 \frac{b}{3a} \; ,
\end{equation}
which leads to a reduction of the generating function
\begin{equation} \label{m1j}
\hat{S}(q',p,E) - q' p = \bar{S} - \varepsilon x - a x^3 -
\frac{\sigma}{2}p'^2+ {\cal O}(|\varepsilon|^{5/2}) \; .
\end{equation}
This is just the usual normal form with the replacement
of $S_0$ by $\bar{S}$. Furthermore the exponential prefactor
in the integral is modified by the Jacobian of the transformation.
After an integration over $p$ we obtain
\begin{equation} \label{m1k}
d_\xi(E) = \mbox{\bf Re} \left[ \frac{
\exp \left( \frac{i}{\hbar} \bar{S} - \frac{i \pi}{2} \nu
- \frac{i \pi}{4} \sigma \right) }{
l \sqrt{2 \pi^3 \hbar^3}} \, 
\int_{-\infty}^\infty \! dx \, \left( T_0 - \frac{2 b}{3 a} T_0 x
- \varepsilon_E x \right)
\exp \left\{ -\frac{i}{\hbar} 
\left( \varepsilon x + a x^3 \right) \right\} \right] \; ,
\end{equation}
where the exponential prefactor has been expanded up to
${\cal O}(x)$. The integral in (\ref{m1k}) can be split into two
terms. The first term with the constant term in the exponential
prefactor has the same form form as the local approximation,
and the second term is proportional to the derivative of the
first term with respect to $\varepsilon$. The integrals can
be found in the section on Airy functions in \cite{AS65}.
The result depends on whether $\varepsilon$ has the
same or the opposite sign as $a$. We first consider the
case when they have opposite sign, i.\,e.\ when the 
orbits are real. The contribution to the level density
is then given by
\begin{eqnarray} \label{m1l}
d_\xi(E) = \mbox{\bf Re} \frac{
\exp \left( \frac{i}{\hbar} \bar{S} - \frac{i \pi}{2} \nu
- \frac{i \pi}{4} \sigma \right) }{
l \sqrt{2 \pi^3 \hbar^3}} \,
\left\{ \frac{2 \pi T_0}{3} 
\sqrt{\left|\frac{\varepsilon}{3 a}\right|}
\left( J_{-1/3} \left(\frac{|\Delta S|}{\hbar}\right)
+      J_{ 1/3} \left(\frac{|\Delta S|}{\hbar}\right) \right)
\right. &&
\nonumber \\ \left.
- \frac{ 2 \pi i \varepsilon}{9 |a|} 
\left( \varepsilon_E + \frac{2 b T_0}{3 a} \right)
\left( J_{-2/3} \left(\frac{|\Delta S|}{\hbar}\right)
-      J_{ 2/3} \left(\frac{|\Delta S|}{\hbar}\right) \right)
\right\} && .
\end{eqnarray}
This result can be expressed also by the Airy function Ai
and its derivative, but it seems more natural to express it
by Bessel functions since then the analogy
to the results for higher repetition numbers is more visible.
All coefficients can be expressed by the classical actions
and amplitudes. The final result is
\begin{eqnarray} \label{m1m}
d_\xi(E) &=& \frac{1}{\pi\hbar} 
\left|\frac{2\pi\Delta S}{3\hbar}\right|^{1/2}
\left\{ \bar{A} \, \cos \left( \frac{\bar S}{\hbar}
- \frac{\pi}{2} \bar{\nu} \right)
\left( J_{-1/3} \left(\frac{|\Delta S|}{\hbar}\right)
+      J_{ 1/3} \left(\frac{|\Delta S|}{\hbar}\right) \right)
\right. \\ \nonumber && \left.
- \mbox{sign}(\Delta S) \, \Delta A \, \cos \left( \frac{\bar S}{\hbar}
- \frac{\pi}{2} (\bar{\nu} - 1) \right)
\left( J_{-2/3} \left(\frac{|\Delta S|}{\hbar}\right)
-      J_{ 2/3} \left(\frac{|\Delta S|}{\hbar}\right) \right) 
\right\} \; .
\end{eqnarray}
Here we replaced $\nu + \sigma/2$ by the mean Maslov index
$\bar{\nu}$. That this is correct can be seen from a stationary
phase evaluation of the integrals which shows the 
$\nu_{1,2} = \nu + \sigma/2 \pm \mbox{sign}(\Delta S)/2$. 
The first term in the curly brackets in (\ref{m1m}) is just the
local approximation in which the stability factors of both orbits
are equal. At the bifurcation ($\varepsilon \rightarrow 0$)
the dependence of the classical quantities on $\varepsilon$
guarantees a finite result. The second term in the curly brackets
is the uniform correction which ensures the correct limit as
$\hbar \rightarrow 0$ for finite $\varepsilon$, since then one
arrives at a sum of two individual contributions for the two orbits,
each of Gutzwiller's type.

On the other side of the bifurcation where the orbits are complex
and $\varepsilon$ and $a$ have the same sign, the evaluation of 
the integral in (\ref{m1k}) leads to
\begin{equation} \label{m1o}
d_\xi(E) = \mbox{\bf Re} \frac{
\exp \left( \frac{i}{\hbar} \bar{S} - \frac{i \pi}{2} \nu
- \frac{i \pi}{4} \sigma \right) }{
l \sqrt{2 \pi^3 \hbar^3}} \,
\left\{ \frac{2 T_0}{3} 
\sqrt{\left|\frac{\varepsilon}{a}\right|}
K_{1/3} \left(\frac{|\Delta S|}{\hbar}\right)
+ \frac{ 2 i \varepsilon}{3 \sqrt{3} |a|} 
\left( \varepsilon_E + \frac{2 b T_0}{3 a} \right)
K_{2/3} \left(\frac{|\Delta S|}{\hbar}\right)
\right\} \; .
\end{equation}
In this equation the mean quantities $\bar{S}$ and $\bar{A}$
are real, but the differences $\Delta S$ and $\Delta A$ are
purely imaginary. The quantities are defined in Eqs.\
(\ref{m1c}), (\ref{m1g}) and (\ref{m1h}) with the
convention for the sign of the square root that has
been discussed above
\begin{equation} \label{m1p}
d_\xi(E) = \mbox{\bf Re \,} \frac{1}{\pi\hbar} 
\left|\frac{2 \Delta S}{\pi \hbar}\right|^{1/2}
\exp  \left( \frac{i}{\hbar} \bar{S}
- \frac{i \pi}{2} \nu - \frac{i \pi}{4} \sigma \right)
\left\{ \bar{A} \, K_{1/3} \left(\frac{|\Delta S|}{\hbar}\right)
+      \Delta A \, K_{2/3} \left(\frac{|\Delta S|}{\hbar}\right)
\right\} \; .
\end{equation}
In the limit that the argument of the $K$-Bessel functions is
large one obtains only the contribution of that orbit for which
the imaginary part of the action is positive
\begin{equation} \label{m1q}
d_\xi(E) = \mbox{\bf Re \,} \frac{A_1}{\pi\hbar} 
\exp  \left( \frac{i}{\hbar} S_1
- \frac{i \pi}{2} \nu - \frac{i \pi}{4} \sigma \right) \; .
\end{equation}
This is connected to Stokes' phenomenon.

\section{Uniform approximation for the period-doubling bifurcation}
\label{secm2}

For the period-doubling bifurcation, we start again with the normal
form $\hat{S}(q',p,E)$ of the action and go beyond the local approximation
by incorporating the next order term in the expansion with respect to $q'$ 
\begin{equation} \label{m2a}
\hat{S}(q',p,E) = S_0 + q'p - \varepsilon q'^2 - a q'^4 - b q'^6
- \frac{\sigma}{2} p^2 \; ,
\end{equation}
where once more $\sigma = \pm 1$.
No term of order $q'^5$ can be present since both satellite fixed
points belong to the same orbit and thus must have the same action.
The fixed points lie at
\begin{eqnarray} \label{m2b}
p_0=0, &\quad& q'_0=0\\
p_1=0, &\quad& q'_1=\pm\sqrt{-\frac{\varepsilon}{2a}}
\left( 1+\frac{3b\varepsilon}{8a^2} \right)
+{\cal O}(|\varepsilon|^{5/2})\;,\nonumber
\end{eqnarray}
where the subscript $0$ indicates the central orbit which has
a repetition number of $2l$, and the subscript $1$ denotes the
satellite orbit. The actions of the orbits are $S_0$ and
\begin{equation} \label{m2c}
S_1 = S_0 + \frac{\varepsilon^2}{4a}
+\frac{b \varepsilon^3}{8 a^3}
+{\cal O}(\varepsilon^4) \; ,
\end{equation}
and the periods are given by $T_0 = \partial \hat{S}/\partial E$ 
evaluated at the origin and
\begin{equation} \label{m2d}
T_1 = T_0 + \frac{\varepsilon_E}{2a} \varepsilon
+{\cal O}(\varepsilon^2) \; .
\end{equation}
The traces of the monodromy matrix follow from (\ref{m1e})
and are given by
\begin{eqnarray} \label{m2e}
\mbox{Tr\,} M_0 &=& 2 - 2 \sigma \varepsilon 
\nonumber \\
\mbox{Tr\,} M_1 &=& 2 + 4 \sigma \varepsilon
- \frac{3 b \sigma}{a^2} \varepsilon^2
+ {\cal O}(\varepsilon^3) \; ,
\end{eqnarray}
and the stability prefactors follow as
\begin{eqnarray} \label{m2f}
A_0 &=& |8 \varepsilon l^2|^{-1/2} T_0 \nonumber\\
A_1 &=& |4 \varepsilon l^2|^{-1/2}
\left( T_0 + \frac{\varepsilon_E}{2 a} \varepsilon
+ \frac{3 b T_0}{8 a^2} \varepsilon
+ {\cal O}(\varepsilon^2) \right) \; ,
\end{eqnarray}
where the repetition numbers of the orbits are $r_0=2l$ and $r_1=l$.

In the following we will use the definitions
\begin{equation} \label{m2g}
\bar{S}  = \frac{S_1 + S_0}{2} \, , \; \; \; \; 
\Delta S = \frac{S_1 - S_0}{2} \, , \; \; \; \; 
\bar{\nu}  = \frac{\nu_1 + \nu_0}{2} \, .
\end{equation}

In order to evaluate the integral (\ref{sec2f}) we want to get rid
of the dependence of the exponent on $q'^6$ and introduce for
this purpose a new variable by 
\begin{equation} \label{m2h}
q'^2 = x^2 - x^4 \frac{b}{2a} \; .
\end{equation}
If one considers $q'$ as a quantity of order 
${\cal O}(|\varepsilon|^{1/2})$ the generating function
reduces to
\begin{equation} \label{m2i}
\hat{S} - q' p = S_0 - \varepsilon x^2 - \tilde{a} x^4  
- \frac{\sigma}{2} p^2 + {\cal O}(\varepsilon^4) \; ,
\end{equation}
where $\tilde{a} = a - b \varepsilon/(2 a)$. One can check that
the new prefactor in front of $x^4$ yields the correct action
(\ref{m2c}) up to order $\varepsilon^3$ at the fixed points.
After integrating over $p$ one arrives at
\begin{equation} \label{m2j}
d_\xi(E) = \mbox{\bf Re} \left[ \frac{
\exp \left( \frac{i}{\hbar} S_0 - \frac{i \pi}{2} \nu
- \frac{i \pi}{4} \sigma \right) }{
l (2 \pi \hbar)^{3/2}} \, 
\int_0^\infty \! dx \, \left( T_0 - \frac{3 b T_0}{4 a} x^2
- \varepsilon_E x^2 \right)
\exp \left\{ -\frac{i}{\hbar} 
\left( \varepsilon x^2 + \tilde{a} x^4 \right) \right\} \right] \; ,
\end{equation}
where the exponential prefactor has been expanded up to order
${\cal O}(x^2)$. Once more the integral splits into two terms
of type
\begin{eqnarray} \label{m2k}
B_1 &=& \int_{-\infty}^{+\infty} dx\,
\exp \left( -\frac{i}{\hbar} (
\varepsilon x^2 - \tilde{a} x^4) \right)
\\ \nonumber
&=& \frac{\pi}{2} \left|\frac{\varepsilon}{2\tilde{a}}\right|^{1/2}
\exp \left\{ \frac{i}{\hbar} \Delta S \right\}
\left(
J_{-1/4} \left(\left|\frac{\Delta S}{\hbar}\right|\right)
e^{-i\sigma_1\pi/8}
- \sigma_1 \tilde\sigma_2
J_{1/4}  \left(\left|\frac{\Delta S}{\hbar}\right|\right)
e^{i\sigma_1\pi/8}
\right)
\end{eqnarray}
and
\begin{eqnarray} \label{m2l}
B_2 &=& \int_{-\infty}^{+\infty} dx \, x^2
\exp \left( -\frac{i}{\hbar} (\varepsilon x^2 + \tilde{a} x^4) \right)
\\ \nonumber
&=& -\frac{\varepsilon}{4\tilde a} B_1 -
\frac{\pi \varepsilon}{8 \tilde a}
\left| \frac{\varepsilon}{2\tilde a} \right|^{1/2}
\exp \left\{ \frac{i}{\hbar} \Delta S \right\}
\left(
J_{3/4}\left(\left|\frac{\Delta S}{\hbar}\right|\right)
e^{i\sigma_1\pi 3/8}
-
\sigma_1\tilde\sigma_2 J_{-3/4}\left(\left|\frac{\Delta
S}{\hbar}\right|\right)
e^{-i\sigma_1\pi 3/8}
\right) \; ,
\end{eqnarray}
where $\Delta S$ as given above, 
$\sigma_1 = \mbox{sign}(\tilde a) = \mbox{sign}(\Delta S)$,
and $\tilde\sigma_2 = \mbox{sign}(\varepsilon)$.
Now we introduce $\sigma_2= -\tilde\sigma_1 \tilde\sigma_2$
to discriminate between both sides of the
bifurcation ($\sigma_2 = 1$ when the satellite orbit is real
and $\sigma_2 = -1$ when it is complex).
Expressing all coefficients by the actions $S_{0,1}$
and stability factors $A_{0,1}$ the final result is
\begin{eqnarray} \label{m2m}
d_\xi(E) &=& \mbox{\bf Re \,} \frac{1}{\pi\hbar}
\left|\frac{\pi \Delta S}{2 \hbar}\right|^{1/2}
\exp \left(\frac{i}{\hbar} \bar{S}
- \frac{i \pi}{2} \nu - \frac{i \pi}{4} \sigma 
\right) \\
&&\times \left\{\left(\frac{A_1}{2}+\frac{A_0}{\sqrt{2}}\right)
\left(\sigma_2 J_{1/4}\left(\frac{|\Delta S|}{\hbar}\right)
e^{i\sigma_1\pi/8}
+J_{-1/4}\left(\frac{|\Delta S|}{\hbar}\right)
e^{-i\sigma_1\pi/8}
\right)\right.
\nonumber\\
&&
\left.
+\left(\frac{A_1}{2}-\frac{A_0}{\sqrt 2}\right)\left(
J_{3/4}\left(\frac{|\Delta S|}{\hbar}\right)
e^{i\sigma_1 3\pi/8}
 +\sigma_2J_{-3/4}\left(\frac{|\Delta S|}{\hbar}\right)
e^{-i\sigma_1 3\pi/8}
\right)
\right\} \;. \nonumber
\end{eqnarray}
In contrast to the tangent bifurcation, all appearing 
classical quantities are real even when the satellite
orbit is complex. In the limit that the argument of the
Bessel functions is large the expression reduces to
a sum of the two Gutzwiller contributions of the 
orbits when $\sigma_2 = 1$, and to the single
Gutzwiller contribution of the central orbit
when $\sigma_2 = -1$.

\section{Uniform approximation for the period-tripling bifurcation}
\label{secm3}

For the period-tripling bifurcation the calculations
are more involved, since the orbits which participate
in the bifurcation lie now in a plane. For the previous
two bifurcations ($m=1$ and $m=2$) the orbits lay on
a line so that one had to treat only
one-dimensional integrals. 

The normal form of the generating function $\hat{S}(q',p,E)$ 
in terms of $q'$- and $p$-coordinates is given by
\begin{equation} \label{m3a}
\hat{S}(p,q',E) = p q' - \frac{\varepsilon}{2} (p^2 + q'^2)
- \frac{a}{2 \sqrt{2}} (p^3 - 3 p q'^2) \; .
\end{equation}
The function $\hat{S}(q',p,E) - q' p$ has four stationary points,
one at the origin which corresponds to the central orbit,
and three others which lie on an equilateral triangle
and correspond to the satellite orbit.
In contrast to the previous two cases, one has to go
two orders beyond the normal form in the expansion of the
generating function $\hat{S}$ since the first correction does not 
change the amplitude prefactor of the satellite orbit. The
higher order terms in this expansion are restricted by
the condition that the three stationary points which
correspond to the satellite orbit have to yield all
three the same action and the same trace of the monodromy
matrix of the satellite orbit. This leads to the expansion
\begin{eqnarray} \label{m3b}
\hat{S}(p,q',E) &=& p q' - \frac{\varepsilon}{2} (p^2 + q'^2)
- \frac{a}{2 \sqrt{2}} (p^3 - 3 p q'^2) 
- \frac{b}{4} (p^4 + 2 p^2 q'^2 + q'^4)
- \frac{9 a^2}{8} p (3 p^2 q' - q'^3)
\nonumber \\ &&
- \frac{c}{4 \sqrt{2}} (p^5 - 2 p^3 q'^2 - 3 p q'^4)
- \frac{27 a^3}{64 \sqrt{2}} (39 p^3 q'^2 - 29 p q'^4)
- \frac{3 a b}{4 \sqrt{2}} (-3 p^4 q' + p^2 q'^3) \; .
\end{eqnarray}
The requirement that the three stationary points yield the
same classical properties allows also a further term
of the form $d q'(p^2 + q'^2)(3 p^2 - q'^2)$, but the
properties of the orbits do not depend on this term
and it can be removed in a later step by modifying
the coefficients (\ref{m3i}) of the transformation 
(\ref{m3h}). For that reason we do not write it.

In the following the expansions of several quantities 
in terms of $\varepsilon$ are given. The action of
the satellite orbit is
\begin{equation} \label{m3d}
S_1 = S_0 - \frac{4}{27 a^2} \varepsilon^3
- \frac{16 b}{81 a^4} \varepsilon^4
- \frac{4(64 b^2 - 243 a^4 - 24 a c)}{729 a^6} \varepsilon^5
+ {\cal O} (\varepsilon^6) \; ,
\end{equation}
and its period is of the form
\begin{equation} \label{m3e}
T_1 = T_0 + \left. \frac{\partial \hat{S}}{\partial E}
\right|_{p=p_1, \, q'=q'_1}
= T_0 - \frac{4 \varepsilon_E}{9 a^2} \varepsilon^2
+ {\cal O} (\varepsilon^3)\;.
\end{equation}

The traces of the stability matrices of the
periodic orbits follow from (\ref{m1e}) and are given by
\begin{eqnarray} \label{m3f}
\mbox{Tr} M_0 &=& 2 - \varepsilon^2
\nonumber \\ 
\mbox{Tr} M_1 
&=& 2 + 3 \varepsilon^2
+ \frac{(2511a^4 - 128 b^2 + 192 a c)}{54 a^4} \varepsilon^4
+ {\cal O} (\varepsilon^5) \; .
\end{eqnarray}
The amplitude prefactors follow as
\begin{eqnarray} \label{m3g}
A_0 &=& \frac{T_0}{3 l \sqrt{|\mbox{Tr} M_0 - 2|}} = 
\frac{T_0}{3 l |\varepsilon|}
\nonumber \\
A_1 &=& \frac{T_1}{l \sqrt{|\mbox{Tr} M_1 - 2|}} = 
\frac{1}{\sqrt{3} l |\varepsilon|} \left(
T_0 - \frac{4 \varepsilon_E}{9 a^2} \varepsilon^2
- \frac{T_0 (2511 a^4 - 128 b^2 + 192 a c)}{324 a^4} 
\varepsilon^2 \right) + {\cal O} (\varepsilon^2) \; ,
\end{eqnarray}
where the repetition numbers of the orbits are $r_0=3l$ and $r_1=l$.

We continue now with the evaluation of the integrals
in (\ref{sec2f}) with the generating function of (\ref{m3b}).
The exponent in the integral
is simplified by applying a transformation of the form
\begin{eqnarray} \label{m3h}
p &=& \p + c_1 \p^2 + c_2 \p \q + c_3 \q^2 + c_4 \p^3
         + c_5 \p^2 \q + c_6 \p \q^2 + c_7 \q^3 
\nonumber \\
q' &=& \q + d_1 \p^2 + d_2 \p \q + d_3 \q^2 + d_4 \p^3
         + d_5 \p^2 \q + d_6 \p \q^2 + d_7 \q^3 \; ,
\end{eqnarray}
which removes the terms proportional to $q'^n p^{4-n}$
and $q'^n p^{5-n}$ in the exponent (within the considered
order of the approximation). The coefficients of the
transformation are given by
\begin{eqnarray} \label{m3i}
c_1 &=& - \frac{b \sqrt{2}}{6 a} - \frac{\sqrt{2}}{216 a^3}
(28 b^2 - 243 a^4 - 24 a c) \varepsilon \; , \; \; \; 
c_2 = - \frac{3 \sqrt{2} a}{4} \; , \; \; \; 
c_3 = -c_1 \; , \; \; \;
\nonumber \\ 
c_4 &=& -\frac{27 a^4 - 2 b^2 + 2 a c}{12 a^2} \; , \; \; \;
c_5 = \frac{b}{4} \; , \; \; \; 
c_6 = - \frac{351 a^4 - 16 b^2 + 16 a c}{96 a^2} \; , \;\;\;
c_7 = - \frac{b}{4} \; ,
\nonumber \\
d_1 &=& \frac{3 \sqrt{2} a}{4} \; , \; \; \; 
d_2 = -2 c_1\; , \; \; \; 
d_3 = 0 \; , \; \; \;
d_4 = -b \; , 
\nonumber \\
d_5 &=& -\frac{27 a^4 - 16 b^2 + 16 a c}{96 a^2} \; , \;\;\;
d_6 = \frac{b}{2} \; , \; \; \; 
d_7 = - \frac{81 a^4 - 8 b^2 + 8 a c}{48 a^2} \; ,
\end{eqnarray}
and the transformation leads to
\begin{eqnarray} \label{m3j}
d_\xi(E) &=& \frac{1}{6 l \pi^2 \hbar^2} \mbox{\bf Re}
\int_{-\infty}^\infty \! d\p \, 
\int_{-\infty}^\infty \! d\q \,
[ \alpha_1 + \alpha_2 (\p^2 + \q^2)] 
\nonumber \\ &&
\cdot \exp\{ \frac{i}{\hbar} \left( S_0 
- \frac{\varepsilon}{2} (\p^2 + \q^2)
- \frac{\tilde{a}}{2 \sqrt{2}} (\p^3 - 3 \p \q^2) \right)
- \frac{i \pi}{2} \nu \} \; ,
\end{eqnarray}
where the exponential prefactor in (\ref{sec2f}) times
the Jacobian of the transformation (\ref{m3h}) has
been expanded up to order $\varepsilon^2$. The
new constants appearing in (\ref{m3j}) are
\begin{eqnarray} \label{m3k}
\alpha_1 &=& T_0 \, , 
\nonumber \\ 
\alpha_2 &=& - \frac{144 \varepsilon_E a^2 + 2511 T_0 a^4 
- 128 T_0 b^2 + 192 T_0 a c}{288 a^2}
= \frac{27 a^2 l}{4 |\varepsilon|} 
\left( \frac{A_1}{2 \sqrt{3}} - \frac{A_0}{2} \right)
\nonumber \\
\tilde{a} &=& a - \frac{2 b}{3 a} \varepsilon
+ \frac{243 a^4 - 28 b^2 + 24 a c}{54 a^3} \varepsilon^2 \; .
\end{eqnarray}
After a change of variables $\p = \sqrt{2I} \cos \Phi$
and $\q = \sqrt{2I} \sin \Phi$ the integrals in (\ref{m3j})
are transformed into 
\begin{eqnarray} \label{m3l}
d_\xi(E) &=& \frac{1}{6 l \pi^2 \hbar^2} \mbox{\bf Re}
\int_0^\infty \! dI \, 
\int_0^{2 \pi} \! d\Phi \,
[ \alpha_1 + 2 \alpha_2 I] \,
\exp \left\{ \frac{i}{\hbar} \left( S_0 
- \varepsilon I - \tilde{a} I^{3/2} \cos(3 \Phi) \right)
- \frac{i \pi}{2} \nu \right\}
\nonumber \\
&=& \frac{1}{3 l \pi \hbar^2} \mbox{\bf Re}
\int_0^\infty \! dI \, 
[ \alpha_1 + 2 \alpha_2 I] \,
\J_0 \left( \frac{\tilde{a} I^{3/2}}{\hbar} \right) \,
\exp \left\{ \frac{i}{\hbar} (S_0 - \varepsilon I) 
- \frac{i \pi}{2} \nu \right\} \; ,
\end{eqnarray}
where the relation
\begin{equation} \label{m3m}
\int_0^{2 \pi} \! d\Phi \, 
\exp\{ i z \cos(m \Phi) \} = 2 \pi \J_0(z)
\end{equation}
has been used. The two remaining integrals are evaluated according
to appendix \ref{secintm3} and result in 
\begin{equation} \label{m3n}
\int_0^\infty \! d I \, 
\J_0 \left( \frac{\tilde{a} I^{3/2}}{\hbar} \right) 
\, \exp \left\{ -\frac{i}{\hbar} \varepsilon I \right\} = 
\frac{\hbar}{|\varepsilon|} \, \sqrt{\frac{2 \pi |\Delta S|}{\hbar}} 
\, \exp \left\{ \frac{i \Delta S}{\hbar} \right\} \,
\left[ \J_{-1/6} 
\left( \frac{|\Delta S|}{\hbar} \right)
+ i \sigma \J_{1/6} 
\left( \frac{|\Delta S|}{\hbar} \right) \right]
\end{equation}
and
\begin{eqnarray} \label{m3o}
\int_0^\infty \! d I \, I \,
\J_0 \left( \frac{\tilde{a} I^{3/2}}{\hbar} \right)
\, \exp \left\{ -\frac{i}{\hbar} \varepsilon I \right\}
&=&
\frac{2 \hbar \varepsilon^2}{9 |\varepsilon| \tilde{a}^2 } \, 
\sqrt{\frac{2 \pi |\Delta S|}{\hbar}} \,
\exp \left\{ \frac{i \Delta S}{\hbar} \right\} \,
\left[ \J_{-1/6} 
\left( \frac{|\Delta S|}{\hbar} \right)
\right. \nonumber \\ && \left.
+ i \sigma \J_{1/6} 
\left( \frac{|\Delta S|}{\hbar} \right) 
+ \J_{-5/6} \left( 
\frac{|\Delta S|}{\hbar} \right)
+ i \sigma \J_{5/6} 
\left( \frac{|\Delta S|}{\hbar} 
\right) \right] \; ,
\end{eqnarray}
where
$\Delta S = (S_1 - S_0)/2 = - 2 \varepsilon^3/(27 \tilde{a}^2)$ 
and $\sigma = \mbox{sign}(\Delta S) = -\mbox{sign}(\varepsilon)$.
Altogether one obtains
\begin{eqnarray} \label{m3p}
d_\xi(E) &=& \frac{1}{\pi \hbar} \, \mbox{\bf Re} \,
\sqrt{\frac{2 \pi |\Delta S|}{\hbar}} \,
\exp \left\{ \frac{i}{\hbar} \bar{S} 
- \frac{i \pi}{2} \nu \right\} \, 
\nonumber \\ &&
\times \left\{ 
\left( \frac{A_0}{2} + \frac{A_1}{2 \sqrt{3}} \right)
\left[ \J_{-1/6} 
\left( \frac{|\Delta S|}{\hbar} \right)
+ i \sigma \J_{1/6} 
\left( \frac{|\Delta S|}{\hbar} \right)
\right] \right.
\nonumber \\ &&
\left. -
\left( \frac{A_0}{2} - \frac{A_1}{2 \sqrt{3}} \right)
\left[ \J_{-5/6} 
\left( \frac{|\Delta S|}{\hbar} \right)
+ i \sigma \J_{5/6} 
\left( \frac{|\Delta S|}{\hbar} \right)
\right] \right\} \; ,
\end{eqnarray}
where $\bar{S}=(S_1+S_0)/2$.

\section{The diffraction integral for the period-tripling bifurcation}
\label{secintm3}

In this section we evaluate the diffraction integral which
appears in the uniform approximation for the period-tripling
bifurcation. We consider first the case $z > 0$:
\begin{eqnarray}
&& \int_0^\infty \! d I \, \J_0 ( I^{3/2} ) \, e^{i z I}
\nonumber \\ \label{int1} &=&
\lim_{\varepsilon \rightarrow 0}
\sum_{n=0}^\infty \frac{(i z)^n}{n!}
\int_0^\infty \! d I \, I^n \, \J_0 ( I^{3/2} ) \, 
e^{- \varepsilon I^{3/2}}
\\ \label{int2} &=&
\lim_{\varepsilon \rightarrow 0} 
\sum_{n=0}^\infty \frac{(i z)^n}{n!} \, \frac{2}{3} \,
(1 + \varepsilon^2)^{-(n+1)/3} \,
\Gamma (\frac{2n}{3} + \frac{2}{3}) \,
\mbox{P}_{(2n-1)/3}
\left( \frac{\varepsilon}{\sqrt{1+\varepsilon^2}} \right)
\\ \label{int3} &=&
\frac{2}{3 \pi} \sum_{n=0}^\infty \frac{(i z)^n}{n!} \,
2^{(2n-1)/3} \,
\Gamma^2 (\frac{n}{3} +\frac{1}{3}) \,
\sin(\frac{\pi n}{3} + \frac{\pi}{3})
\\ \label{int4} &=&
\frac{2}{3} \sum_{n=0}^\infty 
\left( \frac{4 i z^3}{27} \right)^n \,
2^{-1/3} \, \frac{ \Gamma(n + \frac{1}{3})}{
n! \, \Gamma(n + \frac{2}{3})}
+ \frac{2 i z}{3} \sum_{n=0}^\infty 
\left( \frac{4 i z^3}{27} \right)^n \,
2^{1/3} \, \frac{ \Gamma(n + \frac{2}{3})}{
3 \, n! \, \Gamma(n + \frac{4}{3})}
\\ \label{int5} &=&
\frac{2^{2/3} \, \Gamma(\frac{1}{3})}{3 \, \Gamma(\frac{2}{3})}
{}_1\mbox{F}_1 
\left( \frac{1}{3};\frac{2}{3};\frac{4iz^3}{27} \right) 
+ i z \frac{2^{4/3} \, \Gamma(\frac{2}{3})}{9 \, \Gamma(\frac{4}{3})}
{}_1\mbox{F}_1 
\left( \frac{2}{3};\frac{4}{3};\frac{4iz^3}{27} \right) 
\\ \label{int6} &=&
\frac{2}{3} \sqrt{\frac{\pi z}{3}} 
\exp \left( \frac{2iz^3}{27} \right)
\left[ \J_{-1/6} \left( \frac{2z^3}{27} \right)
 + i \, \J_{1/6} \left( \frac{2z^3}{27} \right) \right] \; .
\end{eqnarray}
The parameter $\varepsilon$ has been introduced in (\ref{int1}) 
in order to make the integrals absolutely convergent.
The integrals leading to (\ref{int2}) after a substitution
$x = I^{3/2}$ can be found in \cite{PBM88}. From
(\ref{int2}) to (\ref{int3}) the
limit $\varepsilon \rightarrow 0$ has been performed
and the duplication formula of the Gamma function has
been used. From (\ref{int3}) to (\ref{int4}) the sum has
been split into three parts (by taking every third term, 
respectively) where the third part vanished. Furthermore
the recurrence formula and the triplication formula
of the Gamma function have been used. From
(\ref{int4}) to (\ref{int5}) the definition of 
Kummer's function ${}_1\mbox{F}_1$ has been used, 
and from (\ref{int5}) to (\ref{int6}) the formula 
$\exp(iz) \J_\nu(z) = (z/2)^\nu \, 
{}_1 \mbox{F}_1(\nu+1/2;2\nu+1;2iz) / \Gamma(\nu+1)$ \cite{MOS66}.

The corresponding expression for negative values of $z$
follows from the evenness and oddness of the real and
imaginary part of the integral, respectively. We obtain
\begin{equation} \label{int7}
\int_0^\infty \! d I \, \J_0 (a I^{3/2} ) \, e^{-i z I} = 
\frac{2}{3} \sqrt{\frac{\pi |z|}{3 a^2}} 
\exp\{ - i \frac{2 z^3}{27 a^2} \}
\left[ \J_{-1/6} \left( \frac{2 |z|^3}{27 a^2} \right)
- i \mbox{sign}(z) \J_{1/6} \left( \frac{2 |z|^3}{27 a^2} 
\right) \right] \; ,
\end{equation}
and from the derivative of this integral with respect to $z$
follows
\begin{eqnarray} \label{int8}
\int_0^\infty \! d I \, I \J_0 (a I^{3/2} ) \, e^{-i z I} 
&=& 
\frac{4}{27} \sqrt{\frac{\pi |z|^5}{3 a^6}} 
\exp\{ - i \frac{2 z^3}{27 a^2} \}
\left[ \J_{-1/6} \left( \frac{2 |z|^3}{27 a^2} \right)
- i \mbox{sign}(z) \J_{1/6} \left( \frac{2 |z|^3}{27 a^2} 
\right) \right.
\nonumber \\ &&
\left. + \J_{-5/6} \left( \frac{2 |z|^3}{27 a^2} \right)
- i \mbox{sign}(z) \J_{5/6} \left( \frac{2 |z|^3}{27 a^2} 
\right) \right] \; .
\end{eqnarray}

\newpage

\bibliographystyle{unsrt}
\bibliography{paper}

\begin{thebibliography}{10}

\bibitem{Gut71}
M.~C. Gutzwiller.
\newblock Periodic orbits and classical quantization conditions.
\newblock {\em J. Math. Phys.}, 12:343--358, 1971.

\bibitem{BB72}
R.~Balian and C.~Bloch.
\newblock Distribution of eigenfrequencies for the wave equation in a finite
  domain: {III}. {E}igenfrequency density oscillations.
\newblock {\em Ann. Phys.}, 69:76--160, 1972.

\bibitem{BB74}
R.~Balian and C.~Bloch.
\newblock Solution of the {S}chr\"odinger equation in terms of classical paths.
\newblock {\em Ann. Phys.}, 85:514--545, 1974.

\bibitem{BT76}
M.~V. Berry and M.~Tabor.
\newblock Closed orbits and the regular bound spectrum.
\newblock {\em Proc. R. Soc. Lond. A.}, 349:101--123, 1976.

\bibitem{BT77a}
M.~V. Berry and M.~Tabor.
\newblock Calculating the bound spectrum by path summation in action-angle
  variables.
\newblock {\em J. Phys. A: Math. Gen.}, 10:371--379, 1977.

\bibitem{Gut90}
M.~C. Gutzwiller.
\newblock {\em Chaos in Classical and Quantum Mechanics}.
\newblock Springer, New York, 1990.

\bibitem{CL91}
S.~C. Creagh and R.~G. Littlejohn.
\newblock Semiclassical trace formulas in the presence of continuous
  symmetries.
\newblock {\em Phys. Rev. A}, 44:836--850, 1991.

\bibitem{CL92}
S.~C. Creagh and R.~G. Littlejohn.
\newblock Semiclassical trace formulae for systems with non-{A}belian symmetry.
\newblock {\em J. Phys. A}, 25:1643--1669, 1992.

\bibitem{OH87}
A.~M.~Ozorio de~Almeida and J.~H. Hannay.
\newblock Resonant periodic orbits and the semiclassical energy spectrum.
\newblock {\em J. Phys. A: Math. Gen.}, 20:5873--5883, 1987.

\bibitem{Mey70}
K.~R. Meyer.
\newblock Generic bifurcations of periodic points.
\newblock {\em Trans. Am. Math. Soc.}, 149:95--107, 1970.

\bibitem{Brj70}
A.~D. Brjuno.
\newblock Instability in a {H}amiltonian system and the distribution of
  asteroids.
\newblock {\em Math. USSR Sbornik}, 12:271--312, 1970.

\bibitem{Bru72}
A.~D. Bruno.
\newblock Research on the restricted three body problem. {I}: Periodic
  solutions of a {H}amiltonian system.
\newblock Preprint No. 18, Moskva: Inst. Prikl. Mat. Akad. Nauk SSSR. 44p.
  (Russian), 1972.

\bibitem{Sie96}
M.~Sieber.
\newblock Uniform approximation for bifurcations of periodic orbits with high
  repetition numbers.
\newblock {\em J. Phys. A}, 29:4715--4732, 1996.

\bibitem{Sie96b}
M.~Sieber.
\newblock Semiclassical transition from an elliptical to an oval billiard.
\newblock Preprint ULM-TP/96-4, submitted to J. Phys. A, 1996.

\bibitem{Mil74}
W.~H. Miller.
\newblock Classical-limit quantum mechanics and the theory of molecular
  collisions.
\newblock {\em Adv. Chem. Phys.}, 25:69--177, 1974.

\bibitem{Lit90}
R.~G. Littlejohn.
\newblock Semiclassical structure of trace formulas.
\newblock {\em J. Math. Phys.}, 31:2952--2977, 1990.

\bibitem{SD96}
D.~A. Sadovski\'{\i} and J.~B. Delos.
\newblock Bifurcation of the periodic orbits of {H}amiltonian systems: {A}n
  analysis using normal form theory.
\newblock {\em Phys. Rev. E}, 54:2033--2070, 1996.

\bibitem{BBL92}
E.~Bogomolny, O.~Bohigas, and P.~Leboeuf.
\newblock Distribution of roots of random polynomials.
\newblock {\em Phys. Rev. Lett.}, 68:2726--2729, 1992.

\bibitem{HKS87}
F.~Haake, M.~Ku\'s, and R.~Scharf.
\newblock Classical and quantum chaos for a kicked top.
\newblock {\em Z. Phys. B}, 65:381--395, 1987.

\bibitem{KHE93}
M.~Ku\'s, F.~Haake, and B.~Eckhardt.
\newblock Quantum effects of periodic orbits for the kicked top.
\newblock {\em Z. Phys. B}, 92:221--233, 1993.

\bibitem{BGHS96}
P.~A. Braun, P.~Gerwinski, F.~Haake, and H.~Schomerus.
\newblock Semiclassics of rotation and torsion.
\newblock {\em Z. Phys. B}, 100:115--127, 1996.

\bibitem{KHD93}
M.~Ku\'s, F.~Haake, and D.~Delande.
\newblock Prebifurcation periodic ghost orbits in semiclassical quantization.
\newblock {\em Phys. Rev. Lett.}, 71:2167--2171, 1993.

\bibitem{SSD95}
D.~A. Sadovski\'{\i}, J.~A. Shaw, and J.~B. Delos.
\newblock Organization of sequences of bifurcations of periodic orbits.
\newblock {\em Phys. Rev. Lett.}, 75:2120--2123, 1995.

\bibitem{HS97}
F.~Haake and H.~Schomerus.
\newblock In preparation.

\bibitem{SS97}
H.~Schomerus and M.~Sieber.
\newblock In preparation.

\bibitem{Rim82}
R.~J. Rimmer.
\newblock {\em Generic Bifurcations for Involutory Area Preserving Maps}.
\newblock Memoirs of the AMS No.\ 272. American Mathematical Society,
  Providence, Rhode Island, 1982.

\bibitem{AMBD87}
M.~A. M.~De Aguiar, C.~P. Malta, M.~Baranger, and K.~T.~R. Davies.
\newblock Bifurcations of periodic trajectories in non-integrable {H}amiltonian
  systems with two degrees of freedom: Numerical and analytical results.
\newblock {\em Ann. Phys.}, 180:167--205, 1987.

\bibitem{AS65}
M.~Abramowitz and I.~A. Stegun, editors.
\newblock {\em Handbook of Mathematical Functions}.
\newblock Dover, New York, 1965.

\bibitem{PBM88}
A.~P. Prudnikov, Yu.~A. Brychkov, and O.~I. Marichev.
\newblock {\em Integrals and Series, Volume 2: Special Functions}.
\newblock Gordon and Breach Science Publishers, New York, 1988.
\newblock Second Printing with corrections.

\bibitem{MOS66}
W.~Magnus, F.~Oberhettinger, and R.~P. Soni.
\newblock {\em Formulas and Theorems for the Special Functions of Mathematical
  Physics}.
\newblock Springer, Berlin, 1966.

\end{thebibliography}

\end{document}